\documentclass[12pt]{iopart}
\pdfoutput=1 
\usepackage[pdftex]{graphicx,color}
\usepackage{amssymb}
\usepackage{epsfig}
\usepackage{subfigure}
\usepackage{textcomp}
\usepackage{cite}
\usepackage{float}
\usepackage{bm}
\expandafter\let\csname equation*\endcsname\relax
\expandafter\let\csname endequation*\endcsname\relax
\usepackage{amsmath}
\graphicspath{ {./figs/} }
\def \equi#1{\mathrel{\mathop{\kern 0pt\sim}\limits_{#1}}}

\begin{document}
\title{Starvation Dynamics of a Greedy Forager}

\author{U. Bhat} \address{Department of Physics, Boston University, Boston,
  Massachusetts 02215, USA and Santa Fe Institute, 1399 Hyde Park Road, Santa
  Fe, New Mexico 87501, USA}
  
\author{S. Redner} \address{Santa Fe Institute, 1399 Hyde Park Road, Santa
  Fe, New Mexico 87501, USA}

\author{O. B\'enichou} \address{Laboratoire de Physique Th\'eorique de la
  Mati\`ere Condens\'ee (UMR CNRS 7600), Universit\'e Pierre et Marie Curie,
  4 Place Jussieu, 75252 Paris Cedex France}

\begin{abstract}
  We investigate the dynamics of a greedy forager that moves by random
  walking in an environment where each site initially contains one unit of
  food.  Upon encountering a food-containing site, the forager eats all the
  food there and can subsequently hop an additional $\mathcal{S}$ steps
  without food before starving to death.  Upon encountering an empty site,
  the forager goes hungry and comes one time unit closer to starvation.  We
  investigate the new feature of forager \emph{greed}; if the forager has a
  choice between hopping to an empty site or to a food-containing site in its
  nearest neighborhood, it hops preferentially towards food.  If the
  neighboring sites all contain food or are all empty, the forager hops
  equiprobably to one of these neighbors.  Paradoxically, the lifetime of the
  forager can depend \emph{non-monotonically} on greed, and the sense of the
  non-monotonicity is opposite in one and two dimensions.  Even more
  unexpectedly, the forager lifetime in one dimension is substantially
  enhanced when the greed is \emph{negative}; here the forager tends to avoid
  food in its local neighborhood.  We also determine the average amount of
  food consumed at the instant when the forager starves.  We present
  analytic, heuristic, and numerical results to elucidate these intriguing
  phenomena.

\end{abstract}
\maketitle

\section{Introduction}

A classic approach to account for the phenomenon of foraging is to posit that
a forager has perfect knowledge of its environment and adopts an optimum
strategy to exploit environmental
resources\cite{C76,KR85,SK86,OB90,B91,ASD97,KM01}.  In this class of models,
the motion of the forager is treated in an implicit way, and the rate at
which resources are consumed is specified {\it a priori} as deterministic and
spatially homogeneous~\cite{Oaten:1977,Iwasa:1981,Green:1984}.  A
complementary approach is based on employing a simple or a generalized random
walk to find resources.  Here, the search efficiency is quantified by the
time to reach a specified target.  A wide range of models have been
investigated, including L\'evy walks~\cite{Viswanathan:1999a}, intermittent
walks~\cite{Benichou:2005,Oshanin:2007,Lomholt:2008,Bressloff:2011} and
persistent random walks~\cite{Tejedor:2012}.  Each of these examples have
been shown to minimize this search time under specified conditions.  However,
these models typically do not consider consumption explicitly.

Recently, an alternative description of foraging dynamics was developed in
which the forager has no knowledge of its environment and no
intelligence---the \emph{starving random walk}~\cite{BR14,CBR16}.  In this
model, the forager is unaffected by the presence or absence of food and
always performs an unbiased random walk.  When a forager lands on a
food-containing site, all the food at this site is consumed.  Upon eating, a
forager is fully sated and can subsequently hop $\mathcal{S}$ additional
steps without encountering food before it starves.  However, if the forager
lands on an empty site, the forager goes hungry and comes one time unit
closer to starvation.  Because there is no replenishment of resources, the
local environment of the forager is gradually depleted by consumption.  Thus
its ultimate fate is to starve to death.

Basic questions for this starving random walk are: What is the dependence
of: (a) the total amount of food $\mathcal{N}$ consumed at the instant of
starvation and (b) the average lifetime $\mathcal{T}$ of the forager on
fundamental parameters---the metabolic capacity $\mathcal{S}$ and the spatial
dimension $d$?  It was previously found that $\mathcal{N}$ grows as
$\sqrt{\mathcal{S}}$ for $d=1$ and as $\mathcal{S}^\alpha$ with
$\alpha\approx 1.8$ in the ecologically relevant case of $d=2$.
Correspondingly, the mean lifetime $\mathcal{T}$ grows linearly with
$\mathcal{S}$ for $d=1$, as $\mathcal{S}^\beta$ with $\beta\approx 1.9$ in
$d=2$\footnote{Because the exponents $\alpha$ and $\beta$ are close to 2, it
  is natural to speculate that $\mathcal{N}$ and $\mathcal{T}$ both grow as
  $\mathcal{S}^2$, but modified by logarithmic corrections.  However,
  simulations are unable to distinguish this possibility from a power law
  with exponent 1.8 ($\alpha$) and 1.9 ($\beta$)}, and as
$\exp(\mathcal{S}^\omega)$ for $d\geq 3$, with $\omega\approx 1/2$ in $d=3$,
and with $\omega$ a gradually increasing function of $d$ for
$d>3$~\cite{BR14,CBR16}.  A complete understanding of the dependence of the
food consumed at starvation and the lifetime on $\mathcal{S}$ and $d$ has not
yet been reached.

\section{Model and Preview of Results}

\begin{figure}[ht]
  \center{ \subfigure[~1d]{\includegraphics[width=0.35\textwidth]{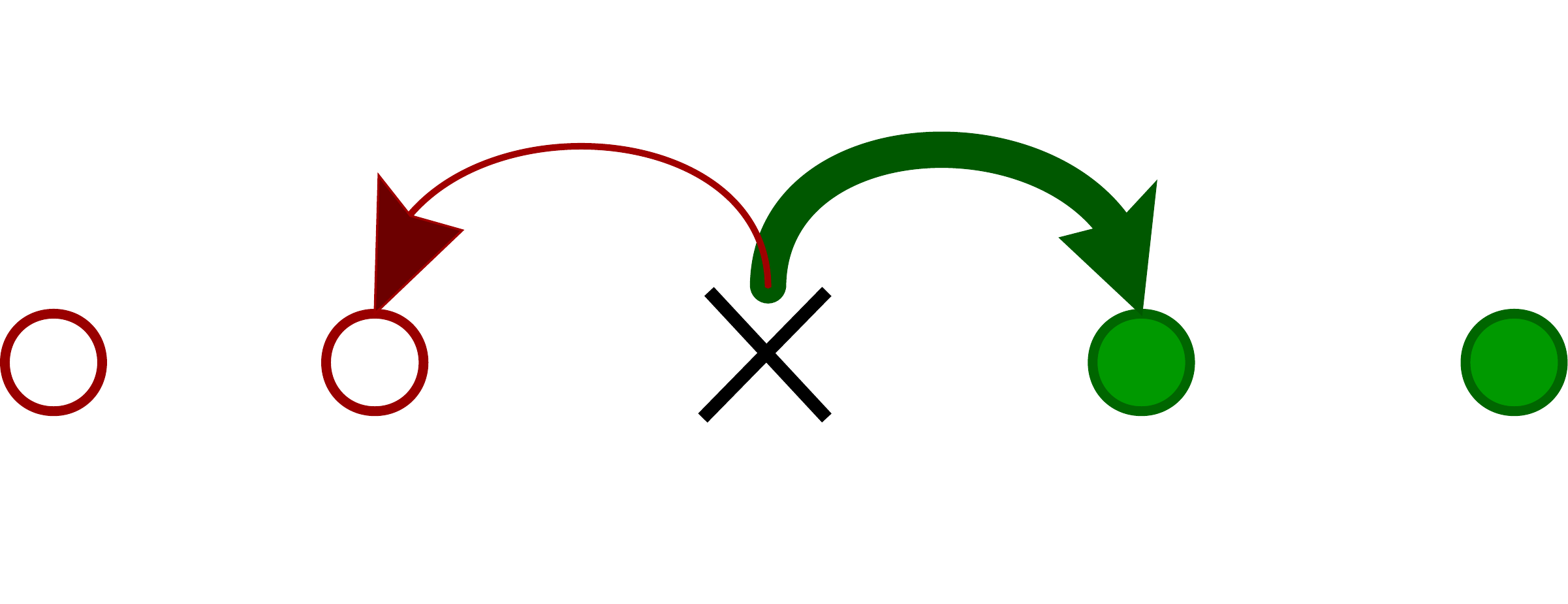}}\hspace{1cm}
    \subfigure[~2d]{\includegraphics[width=0.35\textwidth]{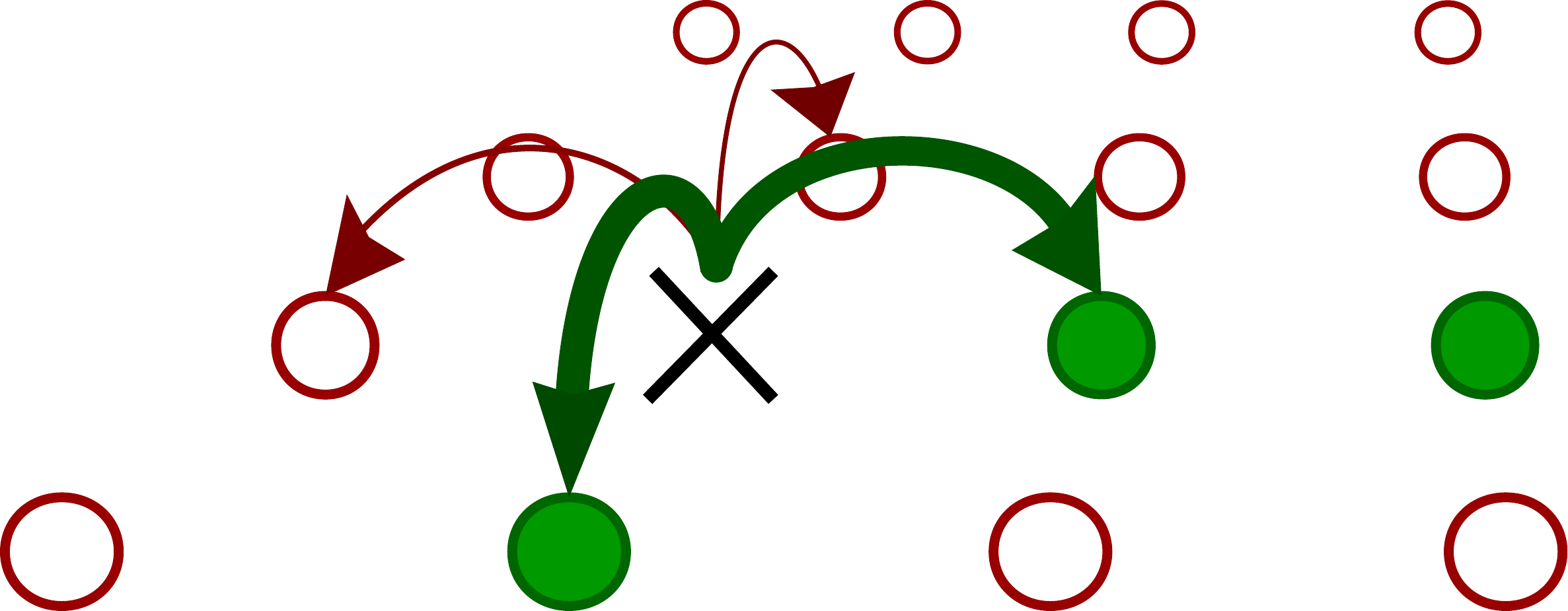}} }
  \caption{ Motion of a greedy forager ($\times$).  Solid and open circles
    indicate food and empty sites, respectively.  Arrow widths indicate
    relative hopping probabilities.}
\label{cartoon}
\end{figure}

In this work, we investigate a natural extension of the starving random walk
to the situation where the forager possesses a minimal level of environmental
awareness at the nearest-neighbor level (a preliminary account of these
results were given in Ref.~\cite{BRB17}).  Namely, whenever the nearest
neighborhood of a forager contains both empty and full (food-containing)
sites, the forager preferentially moves toward one of the full sites
(Fig.~\ref{cartoon}).  We will also investigate the opposite situation in
which the forager tends to avoid food.  We refer to the local propensity to
move towards or away from food as ``greed'', which is quantified by a
greediness parameter $G$, with $-1\leq G\leq 1$.  Positive values of $G$
correspond to a forager that moves preferentially towards food in its nearest
neighborhood, while for $G<0$, the forager tends to avoid food.

In one dimension, we implement greed as follows: when one neighbor of the
forager contains food while the other neighbor is empty, the forager moves
towards the food with probability (Fig.~\ref{cartoon}).
\begin{subequations}
\begin{equation}
\label{p-1d}
p=(1+G)/2\,.
\end{equation}
When the neighboring sites are both empty or both full, then the forager hops
equiprobably to the right or to the left.  For two and higher dimensions,
when there are $k$ neighboring sites that contain food and $z-k$ empty sites
(with $z$ the lattice coordination number), the forager chooses one of
the full sites with probability
\begin{equation}
p={(1+G)}/{\big[(z-k)(1-G)+k(1+G)\big]}\,.
\end{equation}
\end{subequations}

\begin{figure}[ht]
  \center{\subfigure[]{\includegraphics[width=0.325\textwidth]{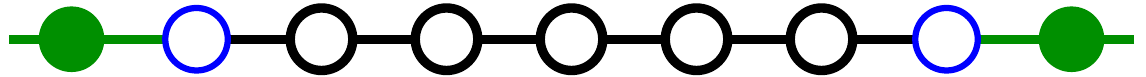}}
\subfigure[]{\includegraphics[width=1.0\textwidth]{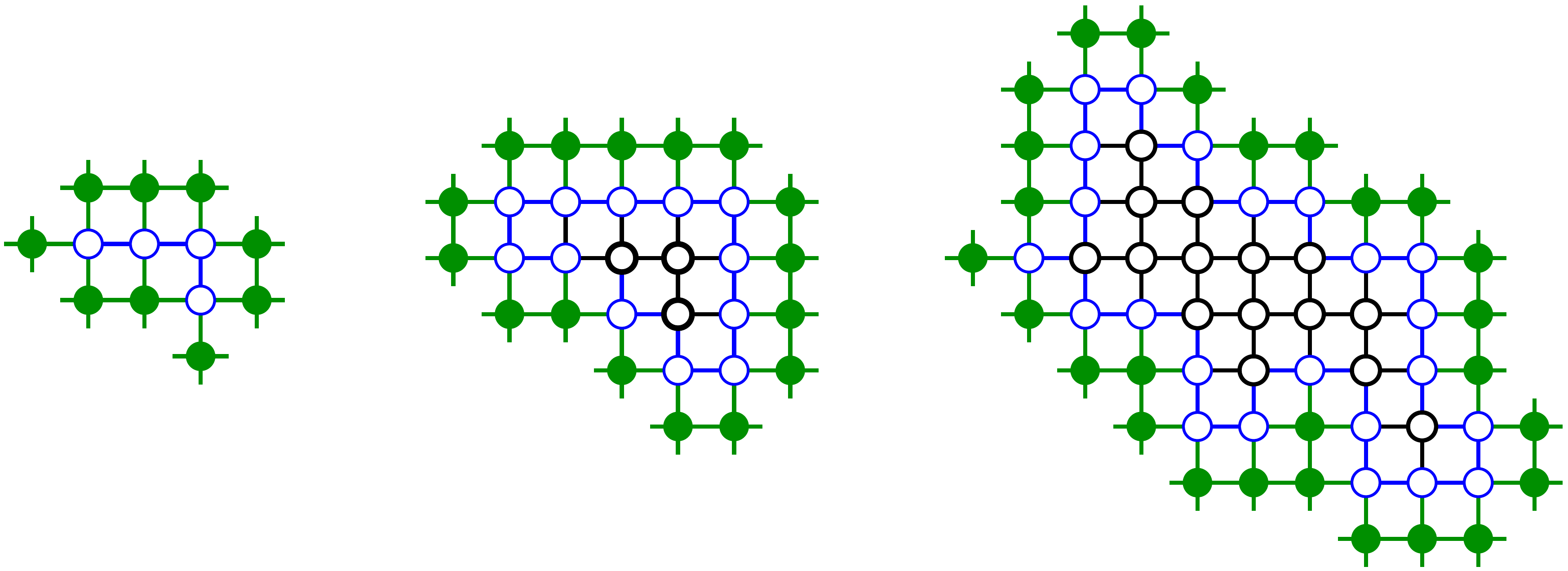}}}
\caption{(a) Illustration of the desert in one dimension and (b) its growth
  in two dimensions.  Black circles: sites in the desert interior, blue
  circles: sites on the desert perimeter, green dots: food.  At
  perimeter sites, the forager is biased to move toward food, for $G>0$, or
  away from food, for $G<0$.}
\label{desert-2d}
\end{figure}

As the walker moves, it creates a depleted region in which food at all sites
along its trajectory has been consumed (Fig.~\ref{desert-2d}).  We term this
region as the ``desert''.  The desert enlarges each time the forager comes to
the perimeter and hops to a food-containing site.  Greed modifies the motion
of the forager only when it is at the desert perimeter.  As the desert grows,
the forager typically spends longer time periods wandering within the desert
without food.  Eventually the forager wanders for $\mathcal{S}$ steps without
encountering food and dies of starvation.

Our implementation of this local greediness represents a particularly simple
feedback between the state of the environment and the forager motion.  Other
classic examples of such feedback include the run and tumble model of
chemotaxis~\cite{BB72,BP77,DZ88}, in which a bacterium effectively swims up a
continuous concentration gradient of nourishment, and infotaxis
models~\cite{EB02,infotaxis}, in which a forager finds a target by moving up
an information gradient.  In classic chemotaxsis models, the concentration of
nutrients is fixed and unaffected by forager consumption.  In contrast, our
modeling explicitly incorporates resource depletion; a related type of
resource depletion also occurs in autochemotaxis~\cite{G05,TKMI16,JKM17}.

\begin{figure}[ht]
  \centerline{
\subfigure[1d]{\includegraphics[width=0.55\textwidth]{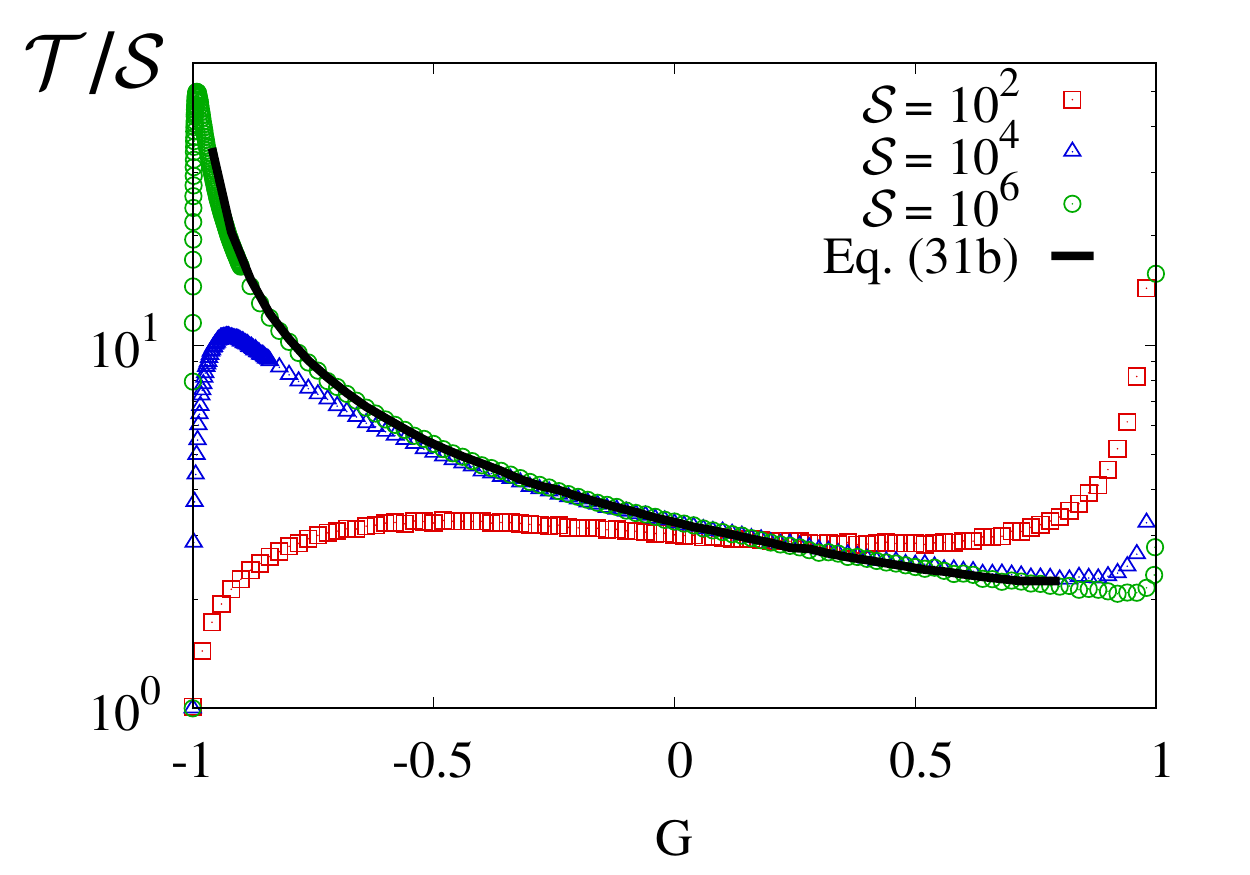}}
\subfigure[2d]{\includegraphics[width=0.55\textwidth]{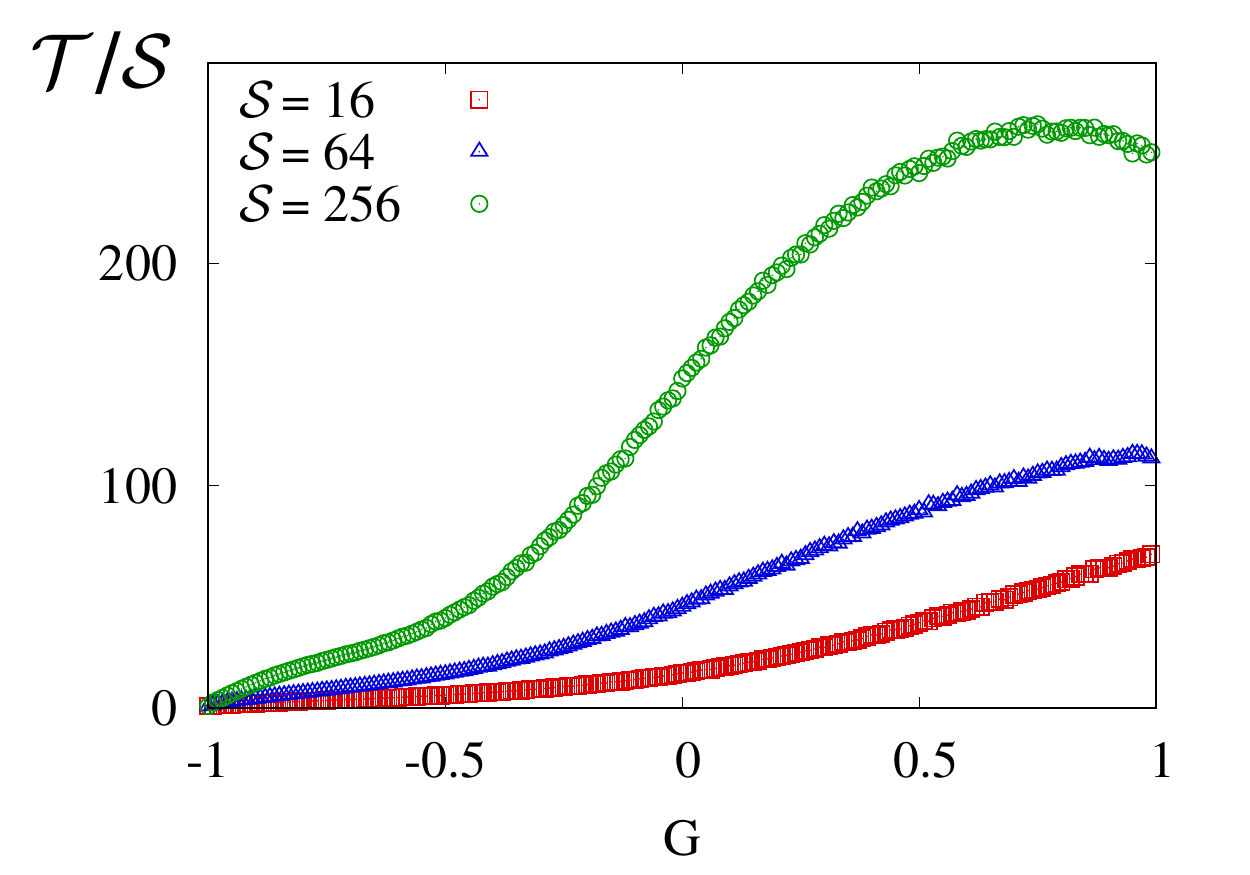}}}
\caption{Our primary result: Simulation data (discrete points) for the
  non-monotonic dependence of the scaled average forager lifetime
  $\mathcal{T}/\mathcal{S}$ on greediness $G=2p-1$ in: (a) one and (b) two
  dimensions.  Error bars are smaller than symbol sizes throughout.  Also
  shown is the analytical prediction for the lifetime; note that the
  numerical evaluation of \eref{T-final} becomes increasingly unstable as
  $G\to 1$.}
\label{greed-prelim}
\end{figure}

Our goal is to determine how greed affects the dependence of the average
forager lifetime $\mathcal{T}$ on its capacity $\mathcal{S}$.  Naively, one
might expect that greed is ``good''\footnote{As quoted by Michael Douglas in
  his role as Gordon Gekko in the movie \emph{Wall Street}.} and always
increases the forager lifetime.  Unexpectedly, we find that greed can be
either good or bad, depending on $\mathcal{S}$ and $d$
(Fig.~\ref{greed-prelim}).  Specifically, the forager lifetime varies
\emph{non-monotonically} with greediness when $\mathcal{S}$ is large, with
opposite senses of the non-monotonicity in $d=1$ and $d=2$. 

While positive greed seems biologically more relevant, the extension to
\emph{negative} greed, where the forager is food averse, leads to surprising
features.  By construction, a forager with negative greed preferentially hops
\emph{away from} food when at the boundary of a desert.  Naively, one might
anticipate that this tendency would decrease the forager lifetime.  However,
for the one-dimensional ``Eden'' initial condition, where all sites initially
contain one unit of food, the forager lifetime grows dramatically as $G$
decreases toward $-1$ (or equivalently $p$ close to 0)
(Fig.~\ref{greed-prelim}(a)).  This growth in the lifetime stops at a
critical value of $G$ that is slightly larger than $-1$; below this point the
lifetime again decreases as $G\to -1$, as it must.

We begin, in Sec.~\ref{sec:fpf}, by deriving general first-passage properties
for a greedy forager in one dimension that moves as a random walk in the
desert interior, but is biased either toward or away from food, when the
forager is at the desert edge.  We then investigate the greedy forager
problem in the simpler case of a one-dimensional semi-infinite geometry
(Sec.~\ref{sec:semi}), where we derive the exact lifetime.  Next, we
investigate the forager lifetime for the Eden initial condition in one
dimension (Sec.~\ref{sec:finite1d}).  We first present a heuristic argument
for the non-monotonic dependence of lifetime on greediness and then develop a
rigorous approach for this dependence.  We also treat the case of a forager
with negative greediness.  In Sec.~\ref{sec:2d} we study greedy forager
dynamics in two dimensions.

\section{First-Passage Formalism}
\label{sec:fpf}

The dynamics of the greedy forager is governed by the probability that it
reaches food in $\mathcal{S}$ steps or less from the last time of its last
meal.  This quantity is just the time integral of the first-passage
probability to reach food up to time $\mathcal{S}$.  Because the motion of a
greedy forager is different at the boundary of the desert than in the
interior, we must extend the first-passage formalism developed in
Refs.~\cite{BR14,CBR16} to account for this boundary perturbation.  In this
section, we summarize these boundary-perturbed first-passage properties for a
greedy forager that will be used in the following sections.
 
Consider a random walk that is either at $x=1$ or $x=L-1$ in the interval
$[0,L]$.  Let $f_L(t)$ denote the probability that an isotropic random walk
\emph{first} reaches either edge of the interval at time step $t$ with this
initial condition.  Throughout, all lengths are expressed in dimensionless
form in units of the lattice spacing $a$.  Now consider a greedy forager at
the edge of the interval.  It hops toward food with probability $p$ and away
from food with probability $1-p$.  In the interior of the interval the
forager hops symmetrically.  We define $F_L(t)$ as the probability that this
greedy random walk, which starts at either $x=1$ or $x=L-1$, reaches either
$x=0$ or $x=L$ at the $t^{\rm th}$ step.  These two first-passage
probabilities are related by the convolution
\begin{equation}
\label{GLbasic}
F_L(t) = p\,\delta_{t,1} + (1-p)\sum_{t'\leq t-1} f_{L-2}(t')\,F_L(t-t'-1)\,.
\end{equation}
The first term accounts for a forager reaching food in a single step.  The
second term accounts for the forager hopping to the interior of the interval.
In the latter case, the walker starts at $x=2$ or $x=L-2$ and hops
symmetrically until it again reaches either $x=1$ or $x=L-1$.  Thus the
relevant first-passage probability is that for an unbiased random walk that
starts at $x=2$ or $x=L-2$ in the interval $[1,L-1]$.  Once the walker first
reaches either $x=1$ or $x=L-1$, the process renews and the subsequent
propagation involves $F_L$.  Since one time unit is used in the first hop to
the right, the walker must reach the boundary in the remaining time $t-t'-1$
steps.

To solve Eq.~\eqref{GLbasic}, we employ the generating functions
\begin{align}
{\widetilde f_L}(z)=\sum_{t\geq 1} f_L(t)\, z^t\,,\qquad\qquad
{\widetilde F_L}(z)=\sum_{t\geq 1} F_L(t)\, z^t\,.\nonumber
\end{align}
to reduce this equation to
${\widetilde F_L}(z) = pz+(1-p)\,z\, {\widetilde f_{L-2}}(z)\,{\widetilde
  F_L}(z)$, with solution
\begin{equation}
\label{gzbasic}
{\widetilde F_L}(z) = \frac{pz}{1-(1-p)\,z \,{\widetilde f_{L-2}}(z)}\,.
\end{equation}

We also exploit first-passage ideas to write formal expressions for two basic
observables: (a) the average amount of food $\mathcal{N}$ consumed when the
forager starves, and (b) the average forager lifetime $\mathcal{T}$.  To
compute these two quantities, we first define the probability $V_k$ that the
forager has visited $k$ distinct sites at the instant of starvation; this is
the same as the probability that the forager has eaten $k$ times at the
instant of starvation.  We can express this probability in the form
\begin{equation}
\label{general_V}
V_k= \big[1-\mathcal{F}_{k}(\mathcal{S})\big]\,\,\prod_{j=1}^{k-1}\mathcal{F}_j(\mathcal{S})\,.
\end{equation}
Here $\mathcal{F}_k(\mathcal{S})$ is the probability that the forager eats
before it starves in a desert of $k$ sites, which can be interpreted as the
probability that the forager successfully escapes a desert of $k$ sites when
it starts one site away from the edge. Thus to create a desert of $k$ sites,
the forager must successively escape a desert of $1,2,3,\ldots,k-1$ sites and
then fail to escape a desert of $k$ sites.  (Note that by definition
$\mathcal{F}_k(\mathcal{S})=1$.)~ In turn, $\mathcal{F}_k(\mathcal{S})$ is
given by
\begin{equation}
\label{general_F}
\mathcal{F}_k(\mathcal{S}) =\sum_{t=0}^\mathcal{S}\,  F_k(t)\,,
\end{equation}
where $F_k(t)$ is the greedy-forager first-passage probability introduced
just above Eq.~\eqref{GLbasic}.  While we tacitly assume a finite interval
length, we will also adapt the formalism above to the case of a semi-infinite
interval in the next section.

The average amount of food consumed by the forager at the instant of
starvation and the average forager lifetime can now be expressed simply in
terms of the distribution of the number of distinct sites visited at
starvation (see also \cite{BR14,CBR16}):
\begin{subequations}
\begin{align}
\label{N-def}
\mathcal{N} &= \sum_{k\geq 0} k V_k\,,\\
\label{general_T}
\mathcal{T}&=\sum_{k\geq 0}\, \big[ \sum_{j=1}^{k-1}\tau_j\big]\,V_k+\mathcal{S}\,,
\end{align}
\end{subequations}
where $\tau_j=\tau_j(\mathcal{S})$ is the average time for a forager to
successfully escape a desert of length $j$ by eating a unit of food at the
desert edge before starvation is reached.  This escape time $\tau_j$ may be
expressed in terms of the first-passage probability $F_j(t)$ of the greedy
forager by
\begin{equation}
\label{general_tau}
 \tau_j=\frac{\sum_{t=0}^\mathcal{S}\,t \,F_j(t)}{\sum_{t=0}^\mathcal{S}\, F_j(t)}\,.
\end{equation}

In the following sections, we will use these general formulae to compute
$\mathcal{N}$ and $\mathcal{T}$ for both the semi-infinite- and finite-desert
geometries.

\section{One Dimension: Semi-Infinite Desert Geometry}
\label{sec:semi}

\subsection{Heuristic Approach}

\begin{figure}[ht]
  \centerline{\includegraphics[width=0.55\textwidth]{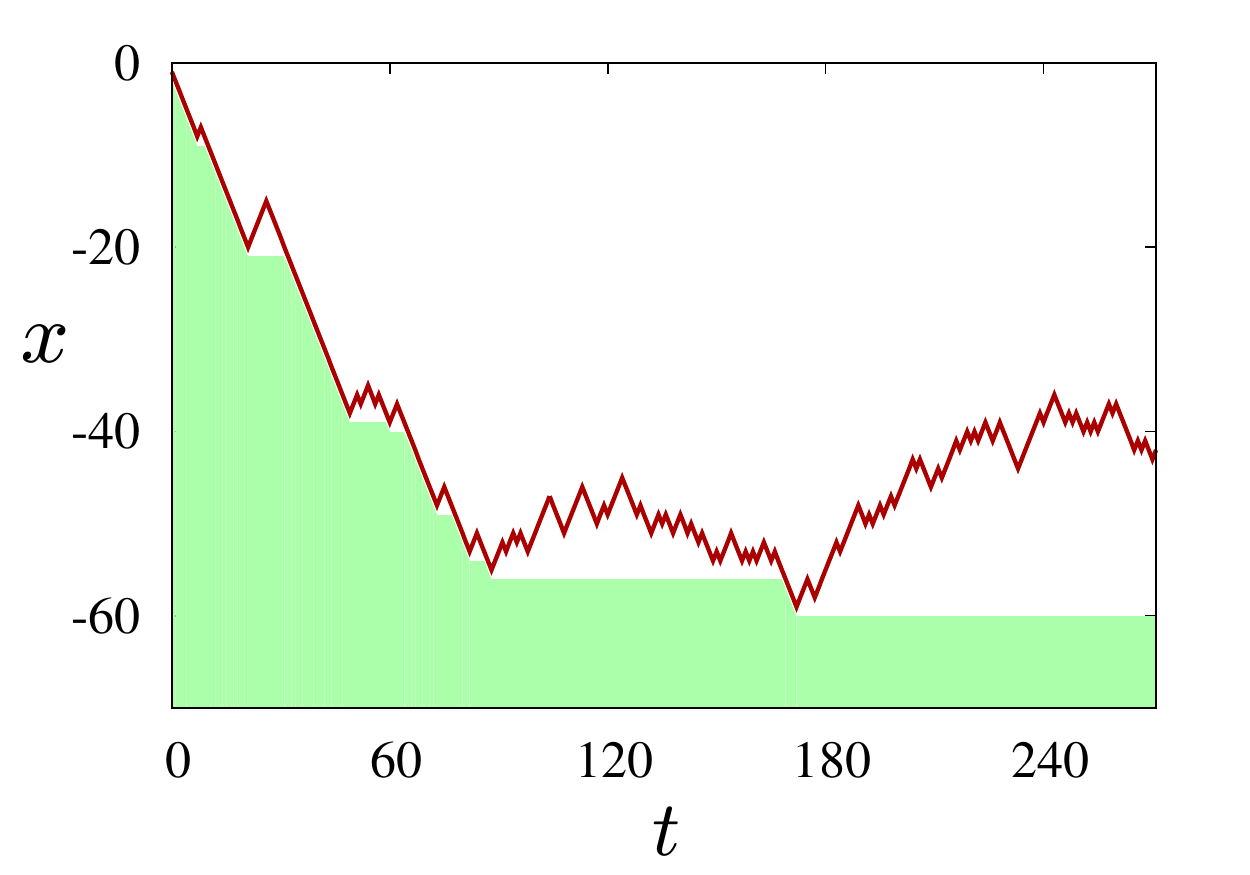}}
  \caption{Space-time trajectory of a greedy forager with lifetime 271 steps
    for $p=0.9$ and $\mathcal{S}=100$ (corresponding to
    $1-p=1/\sqrt{\mathcal{S}}$) in one dimension.  Here, the forager quickly
    carves out a desert whose length precludes reaching the far side at
    $x=0$; thus the far side is irrelevant.}
\label{1d-semi}
\end{figure}

In this geometry, all sites with $x\leq 0$ initially contain food, all sites
with $x\geq 1$ are empty, and the forager begins in a fully sated state at
$x=1$.  When the forager is extremely greedy, corresponding to $p\to 1$, and
also for large $\mathcal{S}$, a typical trajectory consists of segments where
the forager moves ballistically into the food-containing region, interspersed
by diffusive segments in the desert (Fig.~\ref{1d-semi}).  As long as the
diffusive trajectory segment lasts less than $\mathcal{S}$ steps, the forager
returns to the food/desert interface and another cycle of consumption and
subsequent diffusion in the desert begins anew.

We now exploit this picture of alternating ballistic and diffusive segments
to estimate the average forager lifetime.  In a typical trajectory, a
ballistic segment of $m$ consecutive steps towards food followed by a step
away from food occurs with probability $p^m(1-p)$.  The average time for such
a ballistic segment is
\begin{subequations}
\begin{equation}
t_b = \sum_{m\geq 1} m\, p^m\,(1-p)=\frac{p}{1-p}\,.
\end{equation}  
The diffusive segment must return to food within $\mathcal{S}$ steps for the
forager to survive.  Since we are primarily interested in the limit where
$\mathcal{S}$ is large, we may use, without loss of accuracy, continuum
expressions to describe the diffusive segments.  In this continuum limit, the
return probability $\mathcal{R}$ is the integral of the first-passage
probability for a forager with diffusivity $D$ that starts at $x=1$ to reach
$x=0$ within time $\mathcal{S}$~\cite{R01}:
\begin{align}
\mathcal{R}=\int_0^\mathcal{S} dt\,\,\frac{ e^{-1/4Dt}}{\sqrt{4\pi D t^3}} =
\mathrm{erfc}(1/\sqrt{4D\mathcal{S}})\,,\nonumber
\end{align}
where erfc$(\cdot)$ is the complementary error function.  Again in the
continuum limit, the average number of such successful returns is
$\langle r \rangle\!=\! \sum_{r\geq 1} r
\,\mathcal{R}^r(1\!-\!\mathcal{R})\!=\!
\mathcal{R}/(1\!-\!\mathcal{R})\!\simeq\! \sqrt{\pi\mathcal{S}/2}$
for $\mathcal{S}\to\infty$, where the asymptotics of the error function gives
the final result, and we take the diffusion coefficient $D=\frac{1}{2}$ to
correspond to our discrete random-walk simulations.  For a forager that does
return within $\mathcal{S}$ steps, the average return time $t_r$ is
\begin{align}
  t_r  &=\frac{1}{\mathcal{R}} \int_0^\mathcal{S} dt\, t\, \frac{1}{\sqrt{4\pi Dt^3}}\,\,
e^{-1/4Dt}\,,\nonumber\\
&=\sqrt{\frac{2\mathcal{S}}{\pi}}\,
  \frac{e^{-1/2\mathcal{S}}}{\text{erfc}(1/\sqrt{2\mathcal{S}})}
  -1  \simeq \sqrt{\frac{2\mathcal{S}}{\pi}} -1\,.
\end{align}
\end{subequations}
Here and henceforth the symbol $\simeq$ means asymptotically exact as
$\mathcal{S}\to\infty$

\begin{figure}[ht]
\centerline{
\subfigure[]{\includegraphics[width=0.5\textwidth]{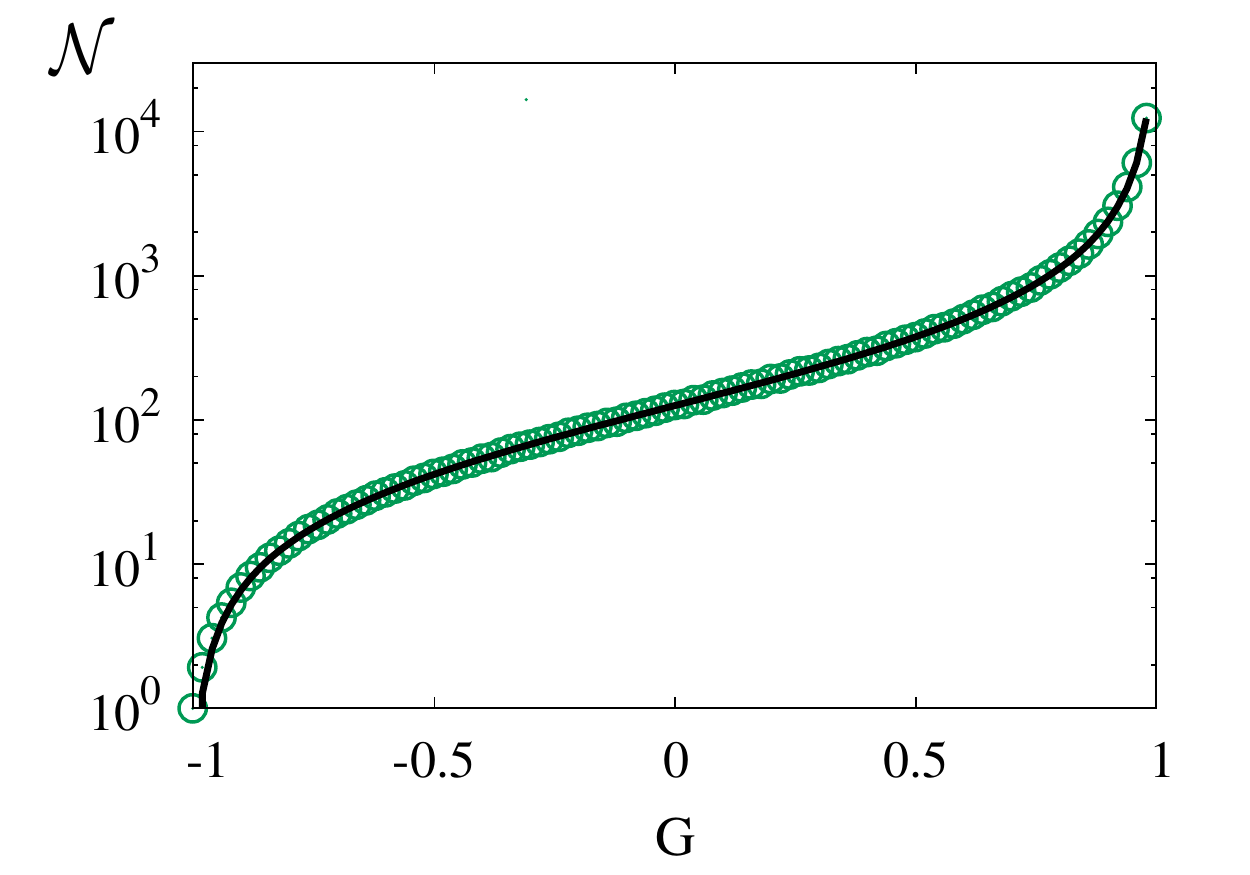}}
\subfigure[]{\includegraphics[width=0.5\textwidth]{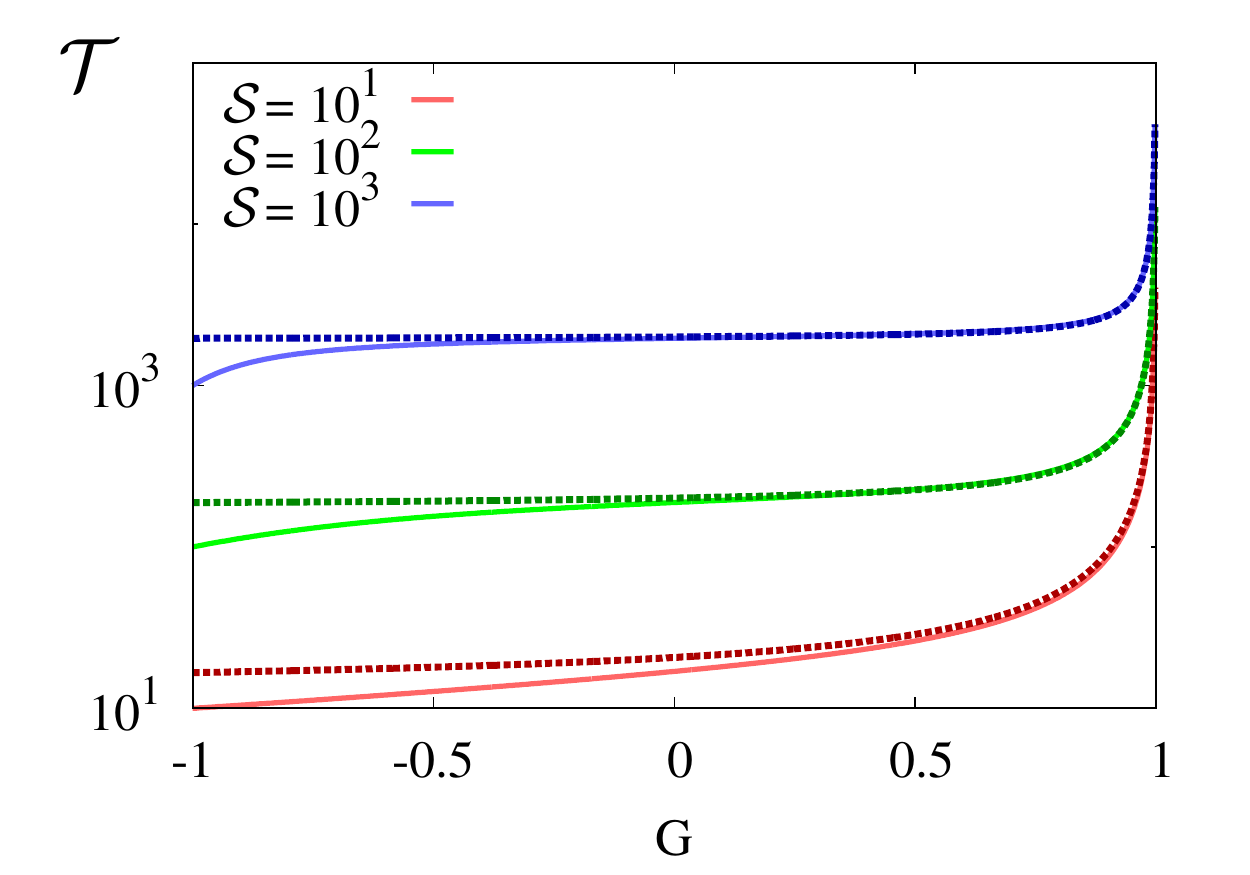}}}
\caption{(a) Average food consumed $\mathcal{N}$ at starvation for the case
  $\mathcal{S}=10^4$, and (b) average forager lifetime $\mathcal{T}$ versus
  greediness $G=2p-1$ for $\mathcal{S}=10^1$, $10^2$, and $10^3$.  The curve
  is the prediction \eqref{n-semi}, while the circles are the simulation
  results.  In (b), the dashed curves are the continuum predictions
  \eqref{T-semi}, while the solid curves are the exact expression
  \eqref{tau-exact}. }
\label{semi-sites}
\end{figure}

The total trajectory therefore consists of
$\langle r\rangle=\sqrt{\pi\mathcal{S}/2}$ elements, each of which are
comprised of a ballistic and a diffusive segment.  Thus the forager eats
$\frac{p}{1-p}$ units of food during each element and the time for each
element equals $t_b+t_r$.  There is also the final and fatal diffusive
segment of exactly $\mathcal{S}$ steps.  Consequently, the food consumed by
the forager and its lifetime, which we respectively write as
$\mathcal{N}_{\rm SI}$ and $\mathcal{T}_{\rm SI}$ (with the subscript SI
denoting the semi-infinite system) respectively, are
(Fig.~\ref{semi-sites})
\begin{subequations}
\begin{align}
\label{n-semi}
\mathcal{N}_{\rm SI}&\simeq \frac{p}{1-p}\,  \sqrt{\frac{\pi\mathcal{S}}{2}}\,,\\
\label{T-semi}
\mathcal{T}_{\rm SI} &\simeq \langle r\rangle (t_b+t_r)+\mathcal{S}=
\frac{2p-1}{1-p}\, \sqrt{\frac{\pi\mathcal{S}}{2}}+2\mathcal{S}\,.
\end{align}
\end{subequations}
For the lifetime, there are two distinct limiting cases:
\begin{itemize}
\item Weak greed ($\frac{1}{1\!-\!p}<\sqrt{\mathcal{S}}$): Lifetime linear in
  $\mathcal{S}$.
\item Strong greed ($\frac{1}{1\!-\!p}>\sqrt{\mathcal{S}}$): Lifetime
  proportional to $\sqrt{\mathcal{S}}$.  However, the amplitude of
  $\sqrt{\mathcal{S}}$ is proportional to $\frac{1}{1-p}$ so that the
  sublinear term is actually larger than $\mathcal{S}$.
\end{itemize}

\subsection{Asymptotic Solution}

We now apply the formalism of Sec.~\ref{sec:fpf} to obtain asymptotic
solutions for $\mathcal{N}_{\rm SI}$ and $\mathcal{T}_{\rm SI}$, as well as
their underlying distributions.  In the semi-infinite geometry, each term in
the product in the distribution of visited sites \eqref{general_V} is
identical because the desert length is always (semi-)infinite.  Thus the
length subscripts for all quantities in Sec.~\ref{sec:fpf} can be dropped.
From \eqref{general_V}, the probability that $k$ units of food have been
eaten at the instant of starvation simplifies to
$V_k=\mathcal{F}^k(\mathcal{S}) \big[1-\mathcal{F}(\mathcal{S})\big]$.  Thus
the average amount of food eaten by the forager at the instant of starvation
is
\begin{equation}
\label{N}
\mathcal{N}_{\rm SI} =\sum_k k V_k = 
\frac{\mathcal{F}(\mathcal{S})}{1-\mathcal{F}(\mathcal{S})}\,.
\end{equation}

To obtain $\mathcal{F}(\mathcal{S})$ we need the underlying first-passage
probability \eqref{gzbasic}.  We start with the well-known expression for the
generating function of the first-passage probability for the isotropic random
walk in the semi-infinite geometry,
${\widetilde f}(z)=(1-\sqrt{1-z^2})/z$~\cite{F68}, and substitute into
Eq.~\eqref{gzbasic} to give
\begin{equation}
\label{ginfty}
{\widetilde F}(z) 
=\frac{z}{1+\frac{1-p}{p}\sqrt{1-z^2}}\,.
\end{equation}
We deduce the long-time behavior of $F(t)$ from the $z\to1$ behavior of\,
$\widetilde{F}(z)$, from which we can compute $\mathcal{F}(\mathcal{S})$, and
finally the amount of food eaten by the forager when it starves,
$\mathcal{N}_{\rm SI}$, and its lifetime $\mathcal{T}_{\rm SI}$.  For large
$\mathcal{S}$, we obtain, both for $p\gg 1/\sqrt{\mathcal{S}}$ and
$p\ll 1/\sqrt{\mathcal{S}}$ (details are given in \ref{app:semi-inf}),
\begin{subequations}
\begin{equation}
\label{eq_N}
\mathcal{N}_{\rm SI} \simeq
\frac{p}{1-p}\sqrt{\frac{\pi\mathcal{S}}{2}}\,,\qquad 
\mathcal{S}\to \infty\,.
\end{equation}

To compute the average forager lifetime, we need the time $\tau_k$ for a
forager to escape a desert of length $k$ (Eq.~\eqref{general_T}).  In the
semi-infinite geometry, all these excursion times are identical,
$\tau_k=\tau$, so that
$\mathcal{T}_{\rm SI} =\tau \mathcal{N}_{\rm SI} + \mathcal{S}$.  The
derivation of $\tau$ is also given in \ref{app:semi-inf} and the final result
for the lifetime has two different forms depending on whether
$p\gg 1/\sqrt{\mathcal{S}}$ or vice versa
\begin{equation}
\label{Tregime1}
\mathcal{T}_{\rm SI}\simeq 
\begin{cases}
\displaystyle{2\mathcal{S}+\frac{(p^2-4p+2)}{p(1-p)}\sqrt{\frac{\pi\mathcal{S}}{2}}}\,,
&\qquad p\gg 1/\sqrt{\mathcal{S}}\,,\\[0.35cm]
\displaystyle{\frac{1}{2} p\mathcal{S}^2+ p\,\sqrt{\frac{\pi}{18}}\,  \mathcal{S}^{3/2}+
\mathcal{S}}\,,&\qquad p\ll 1/\sqrt{\mathcal{S}}\,.
\end{cases}
\end{equation}
\end{subequations}

\subsection{Exact discrete solution}

\begin{figure}[ht]
\centerline{\subfigure[]{\includegraphics[width=0.35\textwidth]{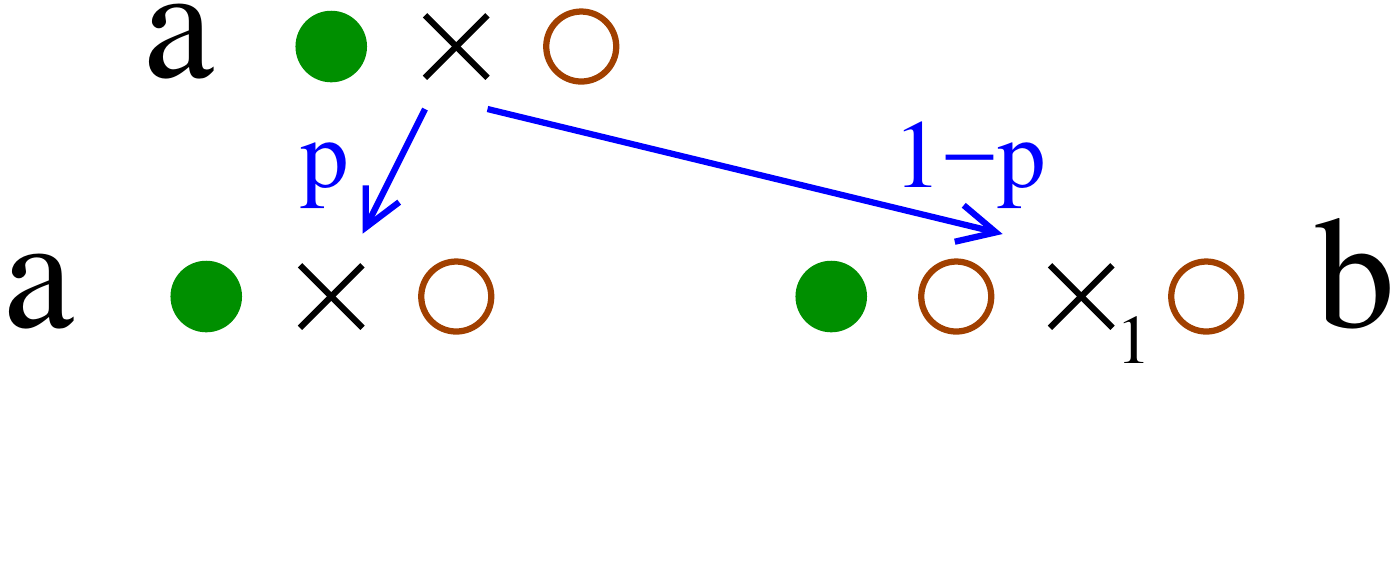}}\qquad\qquad
\subfigure[]{\includegraphics[width=0.45\textwidth]{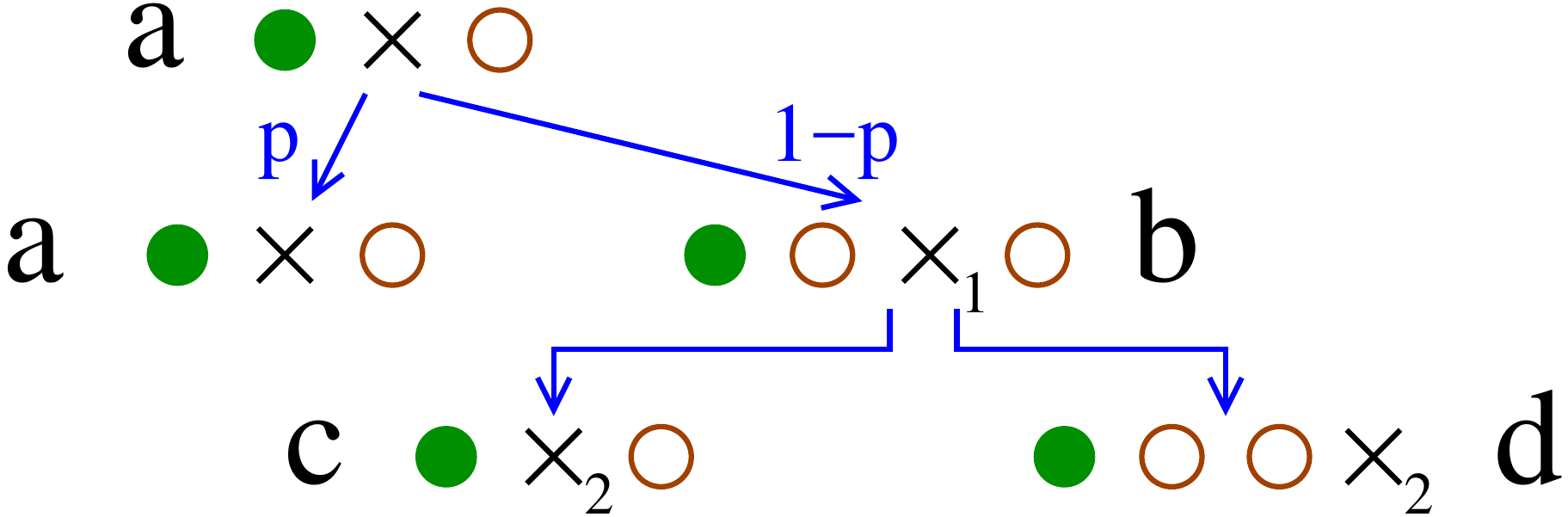}}}
\caption{ (a) State space of the semi-infinite system for $\mathcal{S}=1$.
  Top row: the initial state.  Next row: states after the forager hops once.
  The subscript on the symbol $\times$ denotes the number of time steps that
  the forager is without food.  (b) State space of the semi-infinite system
  for $\mathcal{S}=2$ and 3. }
\label{states-semi-inf}
\end{figure}

We now obtain the exact lifetime for any $p$ and $\mathcal{S}$ by
enumeration.  As a preliminary, first consider small $\mathcal{S}$.  In the
initial state, labeled $a$ in Fig.~\ref{states-semi-inf} the forager is
adjacent to the food.  For $\mathcal{S}=1$, the system can evolve in only two
ways: either the forager moves towards the food, which happens with
probability $p$, or the forager moves away.  In the former case, the forager
eats and the process renews; that is, the system remains in the initial
state.  In the latter case, the forager necessarily dies after one more step.
These evolution steps lead to the state space with two distinct states, $a$
and $b$ (Fig.~\ref{states-semi-inf}(a)).  From this figure, the average
lifetime starting from state $a$ satisfies the backward Kolmogorov
equation~\cite{R01}
\begin{align*}
t_{a} = p(1+t_{a}) + (1-p)\cdot 2 \,,
\end{align*}
which gives
$\mathcal{T}_1\!\equiv\! t_a(\mathcal{S}\!=\!1)\!=\!(2\!-\!p)/(1\!-\!p)$.
This same enumeration can straightforwardly (but more tediously) extended to
larger $\mathcal{S}$.  After one step, the states of the system for
$\mathcal{S}=2$ are identical to those for the case $\mathcal{S}=1$, but the
states after two steps are distinct (Fig.~\ref{states-semi-inf}(b)).  The
corresponding equations for the lifetime starting from any state are:
\begin{align*}
t_{a} &= p\left(1+t_a\right) + (1-p)(t_b+1)\,, \\
t_{b} &= \tfrac{1}{2}(t_c+1)+\tfrac{1}{2}(t_d+1)\,,\\
t_c&=p(t_a+1)+(1-p)\,,\\
t_d&=1\,.
\end{align*}

Solving these equations, the lifetime starting from the initial state $a$ for
$\mathcal{S}=2$ is
\begin{subequations}
\begin{equation}
\mathcal{T}_2\equiv t_a(2)= \frac{2(3-2p)}{(1-p)(2-p)}\,.
\end{equation}
The enumeration for $\mathcal{S}=3$ is the same as that for $\mathcal{S}=2$,
except that the forager lives exactly one more time step before starving.
Thus $\mathcal{T}_3=\mathcal{T}_2+1$.  Following this same approach, the
lifetimes for the next few values of $\mathcal{S}$ are
\begin{align}
  \mathcal{T}_4&= \frac{5(5-5p+p^2)}{(1-p)(8-7p+2p^2)}\,,\qquad  \hspace{1.7cm}\mathcal{T}_5=\mathcal{T}_4+1,\\
  \mathcal{T}_6&= \frac{4(28-35p+13p^2-2p^3)}{(1-p)(16-19p+10p^2-2p^3)}\,,\qquad  \mathcal{T}_7=\mathcal{T}_6+1.
\end{align}
\end{subequations}

We now systematize this enumeration for arbitrary $\mathcal{S}$.  We first
split the set of all trajectories into two categories: (a) set $\textbf{P}$,
which contains all paths that return to food before the forager starves and
(b) set $\textbf{Q}$, which contains all paths where the forager starves.
With this decomposition, the average lifetime for general $\mathcal{S}$ can
be written as
\begin{subequations}
\begin{equation}
\mathcal{T} \, = \, \sum_{\textbf{p} \in \textbf{P}} \mathcal{P}_{\textbf{p}} 
\left( t_\textbf{p} + \mathcal{T} \right) +
\sum_{\textbf{q}\in\textbf{Q}} \mathcal{P}_{\textbf{q}}\, \mathcal{S}\,.
\end{equation}
Here $\mathcal{P}_\textbf{p}$ and $\mathcal{P}_\textbf{q}$ are the
probabilities of paths $\textbf{p}\in\textbf{P}$ and
$\textbf{q}\in\textbf{Q}$, and $t_\textbf{p}$ is the time for the forager to
return to food via path $\textbf{p}$.  The time for all paths in \textbf{Q}
that lead to starvation is simply $\mathcal{S}$.  Rearranging the above
expression gives
\begin{equation}
\mathcal{T} \, = \, \frac{\sum \mathcal{P}_{\textbf{p}} t_{\textbf{p}} 
+ \sum \mathcal{P}_{\textbf{q}} \mathcal{S}}{1 - \sum \mathcal{P}_{\textbf{p}}}\,.
\end{equation}
Since the union  \textbf{P} $\cup$ \textbf{Q} gives all paths, we
have
$\sum_\textbf{p}\! \mathcal{P}_\textbf{p} +
\sum_\textbf{q}\!\mathcal{P}_\textbf{q}=1$.
We use this relation to simplify the second term in the numerator to give
\begin{equation}
\label{T-formal-SI}
\mathcal{T} \, = \, \frac{\sum \mathcal{P}_\textbf{p} t_\textbf{p}}
{1 - \sum \mathcal{P}_\textbf{p}} + \mathcal{S}\,.
\end{equation}
\end{subequations}

\begin{figure}
\centerline{\includegraphics[width=0.5\textwidth]{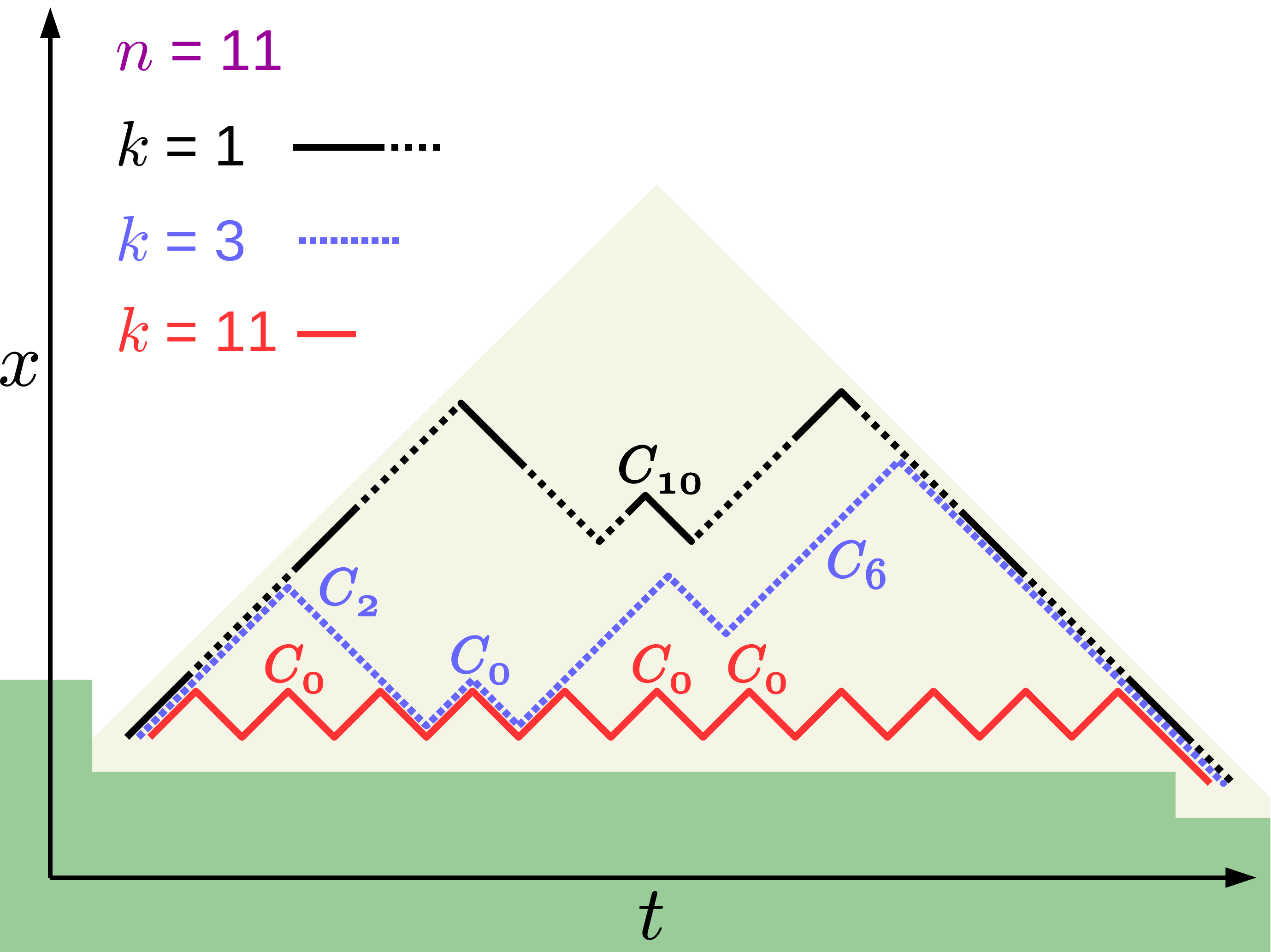}}
\caption{Illustration of paths that reach food in $2n +1 = 23$ steps with
  different $k$, the number of times the path is adjacent to food (green
  strip).  The indicated Catalan numbers show the number of possible paths
  given the constraint of hitting the edge of the desert exactly $k$
  times. The total number of solid-dashed, dotted, and solid paths are
  $\mathcal{C}_{10}$, $\mathcal{C}_2\mathcal{C}_0\mathcal{C}_6$, and
  $(\mathcal{C}_0)^{11}$. }
\label{illust-semi}
\end{figure}

To compute $\sum \mathcal{P}_\textbf{p}$, we first note that the probability
of a particular path depends only on: (i) its length, which we write as
$2n+1$, as a path that starts at $x=1$ can reach food at $x=0$ only in an odd
number of steps, and (ii) the number of times $k$ that the path is adjacent
to food.  The probability of a single path therefore is
\begin{equation*}
  \mathcal{P}_\textbf{p} =  \left(\tfrac{1}{2}\right)^{(2 n-k)}(1-p)^{k}\,p\,.
\end{equation*}
That is, the forager performs an unbiased random walk for the $(2n-k)$ steps
where the forager is not adjacent to food and steps into the desert $k$
times, each with probability $1-p$, when adjacent to food.  In the final
step, the forager reaches food with probability $p$ (Fig.~\ref{illust-semi}).
The sum over all paths that return can be partitioned into sets of paths that
are adjacent to the edge of the desert for exactly $k$ steps.  The number of
paths of this type---of length $2n+1$ with $k$ adjacencies to the desert--is
given
\begin{equation}
\label{cat-def}
 A(n,k) = \frac{(2 n -k -1)!\, k}{(n-k)!\, n!}\,.
\end{equation}
The derivation of this result is given in \ref{app:catalan}.

Using this expression for the number of paths, we have
\begin{equation}
\sum \mathcal{P}_\textbf{p} = p + p\sum_{n=1}^{\lfloor{\mathcal{S}/2}\rfloor} 
\sum_{k=1}^{n} A(n,k) (1-p)^k \left(\tfrac{1}{2}\right)^{(2 n - k)}\,.
\end{equation}
The prefactor $p$ before the sum arises from the last segment of the path in
which the forager consumes one unit of food.  We also write the $n=0$ term
separately as it does not conform to the general expression inside the sum.
Using Eqs.~\eqref{T-formal-SI}--\eqref{cat-def}, the average lifetime for any
$\mathcal{S}$ is given by
\begin{equation}
\label{tau-exact}
\mathcal{T} =  \frac{p +   p\sum_{n=1}^{\lfloor{\mathcal{S}/2}\rfloor} 
\left[(2 n+1)\sum_{k=1}^{n} A(n,k) (1-p)^k
  \left(\frac{1}{2}\right)^{(2 n -
    k)}\right]}{1-p-p\sum_{n=1}^{\lfloor{\mathcal{S}/2}\rfloor}
\left[\sum_{k=1}^{n} A(n,k) (1-p)^k
  \left(\frac{1}{2}\right)^{(2 n - k)}\right]}+ \mathcal{S}\,.
\end{equation}
The comparison between this exact lifetime and the continuum expression
\eqref{Tregime1} is given in Fig.~\ref{semi-sites}(a).  The continuum result
is an excellent approximation for $p>0.8$ for any $\mathcal{S}$.

\section{One Dimension: Finite Desert Geometry}
\label{sec:finite1d}

\begin{figure}[ht]
\centerline{\includegraphics[width=0.6\textwidth]{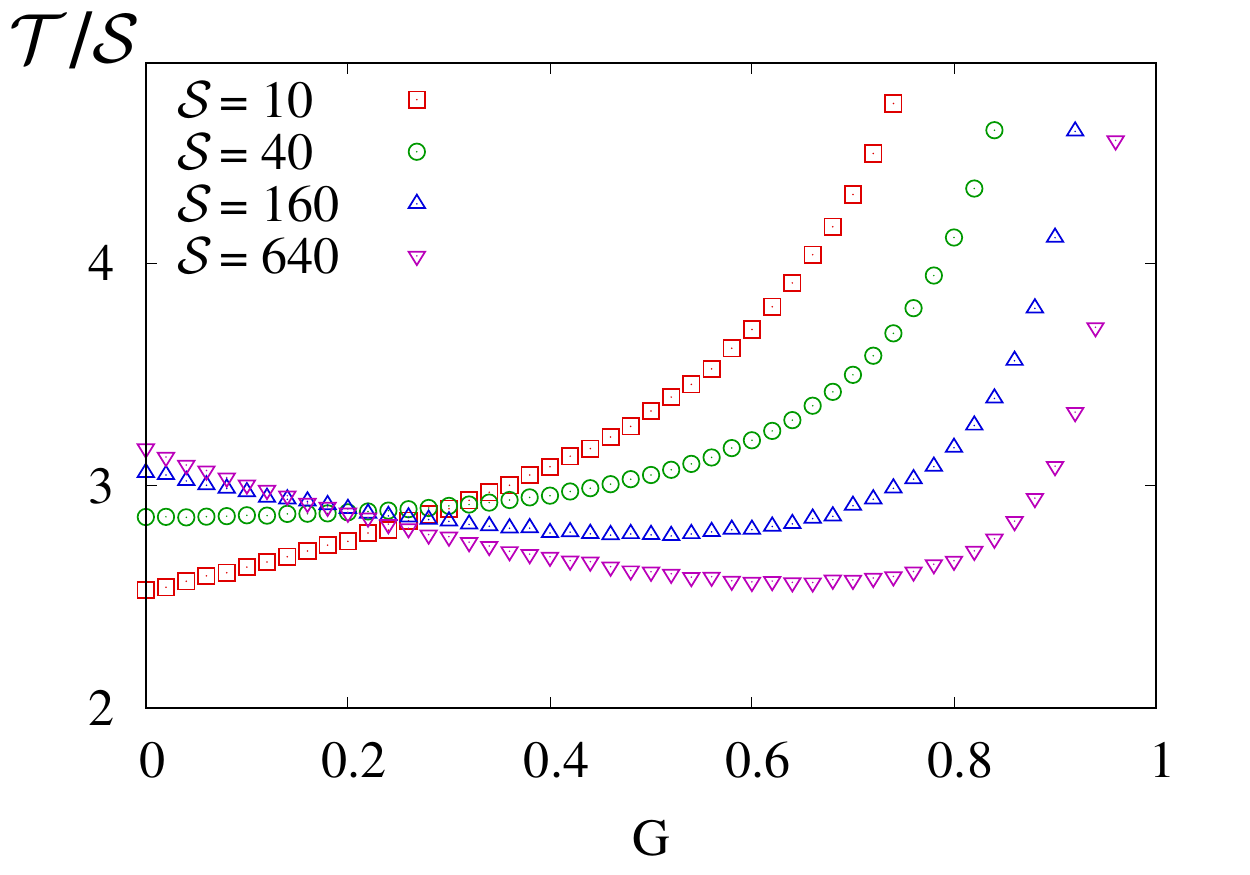}}
\caption{Simulation data for the average forager lifetime $\mathcal{T}$
  versus greediness $G=2p-1$ in one dimension.  The survival times have been
  scaled by $\mathcal{S}$ so that all the data fit onto the same plot.}
\label{tau-1d}
\end{figure}

We now turn to the geometry where each site initially contains food---the
Eden initial condition---and the forager gradually carves out a
finite-length desert.  Unexpectedly, the average forager lifetime varies
\emph{non-monotonically} with (positive) greediness when
$\mathcal{S}>\mathcal{S}^*$, with $\mathcal{S}^*\approx 45$
(Fig.~\ref{tau-1d})---a little greed is bad for a sufficiently ``rich''
forager, but extreme greed is good.

\subsection{Heuristics}
\label{subsec:heuristic}

We first present a heuristic argument for both the food consumed at
starvation $\mathcal{N}$ and the lifetime $\mathcal{T}$.  Our argument
predicts both the non-monotonic lifetime for positive greediness and a huge
maximum in the lifetime for a negative greediness
(Fig.~\ref{greed-prelim}(a)).  In our approach, starvation proceeds in two
stages\footnote{This argument represents both an extension and a
  simplification of the intuitive picture for the starvation process given in
  Ref.~\cite{CBR16}}:

\begin{enumerate}
\item The forager first carves a critical-length desert by repeatedly
  reaching either edge of the desert within $\mathcal{S}$ time steps after
  food is consumed.  The critical length $L_c$ is such that a forager of
  capacity $\mathcal{S}$ typically starves when it attempts to cross a desert
  of length $L>L_c$.  We define the time to create a critical-length desert
  as $\mathcal{T}_{(\text{i})}$ and the food consumed in this phase as
  $\mathcal{N}_{(\text{i})}$.

\item Once the desert length reaches $L_c$, the forager likely starves if it
  attempts to cross the desert.  That is, the far side is unreachable and
  thus is irrelevant.  The time for this second stage,
  $\mathcal{T}_{(\text{ii})}$ is therefore just the mean lifetime
  $\mathcal{T}_{\rm SI}$ in a semi-infinite desert.  Similarly, the amount of
  food consumed in this phase is $\mathcal{N}_{\rm SI}$.
\end{enumerate}

To adapt the above argument to a greedy forager, we need the exit
probabilities and exit times in a finite interval for the random walk that
mimics the motion of the greedy forager---isotropic hopping in the desert
interior and hopping towards food with probability $p$ and away from food
with probability $1-p$ at the desert edge; these quantities are derived in
\ref{sec:escape}.  When the forager starts a unit distance from food in a
desert of length $L$, the average time to reach food is (Eq.~\ref{tn:app})
\begin{equation}
\label{t1}
t_1=\frac{1-p}{p}\, L + 3 - \frac{2}{p}\,.
\end{equation}
Notice that when $1-p\ll \frac{1}{L}$, the mean time to reach food when
starting at $n=1$ approaches 1, while for no greed, $t_1=L-1$.
By definition
\begin{subequations}
\begin{equation}
\mathcal{N}_{(\text{i})}=L_c\,,
\end{equation}
while from \eqref{t1}, the time to reach the critical length $L_c$ is
\begin{equation}
  \mathcal{T}_{(\text{i})}=\frac{1-p}{p}\,\frac{L_c(L_c+1)}{2} + \Big(3 -
  \frac{2}{p}\Big)L_c
\simeq \frac{1-p}{p}\,\frac{L_c^2}{2}\,.
\end{equation}
\end{subequations}
For large $\mathcal{S}$ and hence large $L_c$, we have
\begin{align}
\begin{split}
\label{heuristicsT}
  \mathcal{N}& =  \mathcal{N}_{(\text{i})}+  \mathcal{N}_{(\text{ii})}
=L_c+\mathcal{N}_{\rm SI}\,,\\
\mathcal{T}&=  \mathcal{T}_{(\text{i})}+ \mathcal{T}_{(\text{ii})} 
\simeq\frac{1-p}{p}\frac{L_c^2}{2} + \mathcal{T}_{\rm SI}\,.
\end{split}
\end{align}

We now use the crossing time from Eq.~\eqref{app:tx} for an interval of length $L\gg 1$
\begin{equation}
\label{tmL-1}
t_{\times}\approx \frac{2}{3}\,L^2 + \frac{4}{3}\,\frac{L}{p}\,,
\end{equation}
and set $t_\times$ equal to $\mathcal{S}$ to give the critical desert
length
\begin{equation}
\label{Lc}
L_c\simeq
\begin{cases}
  \sqrt{3\mathcal{S}/2} &\qquad p\gg 1/\sqrt{\mathcal{S}}\,,\\
  ~3p\mathcal{S}/4&\qquad p\ll 1/\sqrt{\mathcal{S}}\,.
\end{cases}
\end{equation}
We emphasize that different behaviors arise for $p\gg 1/\sqrt{\mathcal{S}}$
and $p\ll 1/\sqrt{\mathcal{S}}$.
Using the expression for $\mathcal{T}_{\rm SI}$ in \eqref{Tregime1} together
with \eqref{heuristicsT} and \eqref{Lc}, we have, for
$p\gg1/\sqrt{\mathcal{S}}$,
\begin{subequations}
\begin{align}
\mathcal{N} &\simeq
              \sqrt{\frac{3\mathcal{S}}{2}}
+\frac{p}{1-p}\,\sqrt{\frac{\pi\mathcal{S}}{2}}\,,\\
\label{T_qualit}
\mathcal{T} &\simeq \frac{1-p}{p}\,\frac{3\mathcal{S}}{4}+  2\mathcal{S} 
-\frac{(p^2-4p+2)}{p(1-p)}\,\sqrt{\frac{\pi\mathcal{S}}{2}}\,,
\end{align}
\end{subequations}
while for $p\ll1/\sqrt{\mathcal{S}}$
\begin{subequations}
\begin{align}
\label{N_qualit2}
\mathcal{N} &\simeq \frac{3}{4}\,p\mathcal{S}
+\frac{p}{1-p}\,\sqrt{\frac{\pi\mathcal{S}}{2}}\,,\\
\label{T_qualit2}
\mathcal{T} &\simeq \frac{9}{32}\,p\mathcal{S}^2+ \mathcal{S} 
+  p\,\sqrt{\frac{\pi}{18}}\,\,  \mathcal{S}^{3/2}\,.
\end{align}
\end{subequations}

Two important consequences follow from the above expressions for
$\mathcal{T}$, as illustrated in Figs.~\ref{greed-prelim} and \ref{tau-1d}:
\begin{itemize}
\item Expanding Eq.~\eqref{T_qualit} for $p=\frac{1}{2}+\epsilon$ with
  $\epsilon\to0$ gives
\begin{equation*}
  \mathcal{T} \simeq\frac{11}{4}\mathcal{S}+\Big(12
  \sqrt{\frac{\pi\mathcal{S}}{2}} -10 \mathcal{S}  \Big)\epsilon\,.
\end{equation*}
Thus the lifetime initially increases with $\epsilon$ for
$\mathcal{S}< \mathcal{S}^*={18\pi}/{25}$ and initially decreases otherwise.
While the numerical value of $\mathcal{S}^*$ should not be taken seriously
because of the crudeness of our argument, the important point is that
$\mathcal{S}$ is a non-monotonic function of greediness for
$\mathcal{S}> \mathcal{S}^*$ because the lifetime must eventually increase
with greediness as $G\to 1$.

\item At the crossover, where $p\sim 1/\sqrt{\mathcal{S}}$,
  Eqs.~\eqref{T_qualit} and \eqref{T_qualit2} give the common lifetime
  $\mathcal{T}\sim \mathcal{S}^{3/2}$.  A huge maximum!  This maximum arises
  because the forager eats only when it absolutely must.  Because desert is
  small for $G\approx -1$ (see Eq.~\eqref{N_qualit2}), the strategy of
  avoiding food until nearly $\mathcal{S}$ steps have elapsed is not that
  risky.
\end{itemize}

\subsection{First-passage approach}

We now determine the amount of food consumed at starvation, $\mathcal{N}$,
and the lifetime $\mathcal{T}$ of a greedy forager by extending the approach
of Refs.~\cite{BR14,CBR16} to account for greed.  We start with the
first-passage probability for pure diffusion in the interval $[0,L]$ and then
compute the first-passage probability for greedy forager motion in this same
interval.  From this, we obtain the probability $\mathcal{F}_L(\mathcal{S})$
that the greedy forager can escape a desert of length $L$, as well as the
escape time time $\tau_L$ for this event.  From these two quantities we
finally determine $\mathcal{N}$ and $\mathcal{T}$.

The Laplace transform of the first-passage probability for  diffusion on
$[0,L]$ is~\cite{R01}
\begin{align}
{\widetilde f_L}(s)=
\frac{\sinh\left(\sqrt{\frac{s}{D}}\right)+
\sinh\left(\sqrt{\frac{s}{D}}(L-1)\right)}{\cosh\sqrt{\frac{s}{D}}L}
&=\cosh\sqrt{\frac{s}{D}}-\tanh\sqrt{\frac{sL^2}{4D}}\,\sinh\sqrt{\frac{s}{D}}\,,\nonumber\\
&\underset{s\to 0}{\longrightarrow}
 1-\sqrt{\frac{s}{D}}\tanh\sqrt{\frac{sL^2}{4D}}+\cdots\,.\nonumber
\end{align}
Substituting this expression in \eqref{gzbasic} and converting the generating
function to a continuum Laplace transform by replacing $z\to 1-s$, the
Laplace transform of the first-passage probability for greedy forager motion
for $s\to 0$ and $L\to\infty$ is
\begin{equation}
\label{finite1}
{\widetilde F_L}(s)
=\bigg(1+\frac{1-p}{p}\sqrt{\frac{s}{D}}\tanh\sqrt{\frac{sL^2}{4D}}\bigg)^{-1}.
\end{equation}
Note that this expression reproduces the discrete generating function
${\widetilde F}(z)$ in Eq.~\eqref{ginfty} for $L\to\infty$ with $z\to 1-s$
and $D=1/2$.  As in the semi-infinite geometry, we must separately examine
the limits $p\gg 1/\sqrt{\mathcal{S}}$ and $p\ll 1/\sqrt{\mathcal{S}}$.

\subsection*{The regime $p\gg 1/\sqrt{\mathcal{S}}$:}

Since the Laplace variable $s$ corresponds to $1/\mathcal{S}$ for large
$\mathcal{S}$, the limit $p\gg 1/\sqrt{\mathcal{S}}$ corresponds to
$\sqrt{s}/p\ll 1$.  In this case Eq.~\eqref{finite1} simplifies to
\begin{equation}
\label{regular1}
{\widetilde F_L}(s)\simeq 1-\frac{1-p}{p}\sqrt{\frac{s}{D}}
\tanh\sqrt{\frac{sL^2}{4D}}+\cdots\,.
\end{equation}
Since $\mathcal{F}_L(\mathcal{S}) = \int_0^{\mathcal{S}} F_L(t)\,dt$ (see
Eq.~\eqref{general_F}), their Laplace transforms are related by
$\widetilde{\mathcal{F}_L}(s)={\widetilde F_L}(s)/s$~\cite{AWH13}.  Taking
the inverse Laplace transform, the probability that a greedy forager escapes
an interval of length $L$ within $\mathcal{S}$ steps is
\begin{subequations}
\begin{equation}
\mathcal{F}_L(\mathcal{S})\simeq 1-\frac{1-p}{p}\frac{1}{2\pi i \sqrt{D}}
\int_C {\rm d}s\,\, e^{s\,\mathcal{S}}\,\frac{1}{\sqrt{s}}\tanh \sqrt{\frac{sL^2}{4D}}\,,
\end{equation}
where the vertical segment of the Bromwich contour $C$ lies to the right of
all poles of the integrand.  These poles are located at
$s_n=-D\pi^2(2n+1)^2/L^2$, with $n\in{\mathbb N}$, so that
\begin{equation}
\label{calFk}
\mathcal{F}_L(\mathcal{S})\simeq 1-\frac{1-p}{p}\,\frac{4}{L}\sum_{n=0}^\infty 
e^{-D\mathcal{S}\left[\pi (2n+1)\right]^2/L^2}.
\end{equation}
\end{subequations}
The above represents the exact Laplace inversion of the asymptotic form for
$\widetilde{\mathcal{F}_L}(s)$.

To compute the time $\tau_L$ for a greedy forager to escape an interval of
length $L$, we use the fact that for large $\mathcal{S}$ we can replace the
denominator in the definition~\eqref{general_tau} for $\tau_L$, which is the
probability for the forager to escape an interval of length $L$ within a time
$\mathcal{S}$, by 1. We then use standard Laplace transform manipulations to
give, for large $\mathcal{S}$,
\begin{equation}
\label{LT}
{\widetilde \tau_L}(s)
\simeq -	\frac{1}{s} \frac{\partial}{\partial s}{\widetilde F_L} (s)\,.
\end{equation}
Since the $s$-dependent term in Eq.~\eqref{regular1} for
${\widetilde F_L}(s)$ has the prefactor $(1-p)/p$, we have the general
relation
\begin{equation}
{\widetilde \tau_L}(s)\simeq \frac{1-p}{p}\,\,{\widetilde \tau_L}(s; p=\tfrac{1}{2})
\end{equation}
between the escape time for the greedy forager and for a pure random-walk
forager.  Here, the right-hand side is just the escape time for the case
$p=1/2$ (random walk).  Thus we obtain the fundamental relation between the
escape times
\begin{align}
\label{tauregime1}
  \tau_L&\simeq \frac{1\!-\!p}{p}\,\tau_L(p=\tfrac{1}{2})
        =\frac{1\!-\!p}{p}\int_0^\theta \!\! d u \, u \sum_{j=0}^\infty\frac{4}{(2j\!+\!1)^2}\left\{\!
          1-e^{-(2j+1)^2/u^2}\left[1+\left(\!\frac{2j\!+\!1}{u}\!\right)^2\right]\!\right\}\,.
\end{align}
For the second line, we copy the expression for $\tau_L$ in \cite{BR14,CBR16}
for the case of no greed, and we express the final result in terms of the
natural scaling variable $\theta = L/(\pi\sqrt{D\mathcal{S}})$.

We now use these results for $\mathcal{F}_L$ and $\tau_L$ to determine
$\mathcal{N}$ and $\mathcal{T}$ (see \ref{app:NT} for details):
\begin{subequations}
\begin{equation}
\label{N-final}
\mathcal{N} =\mathcal{N}^*\,\,\frac{4(1-p)}{p}
\int_0^\infty {\rm d}\theta\exp\left[-\frac{2(1-p)}{p}
\sum_{n\ge0}E_1\left(\frac{(2n+1)^2}{\theta^2}\right)\right]
\sum_{n\ge0}e^{-(2n+1)^2/\theta^2},
\end{equation}
where $\mathcal{N}^*\equiv \pi \sqrt{D\mathcal{S}}$ and
$E_1(x)=\int_1^\infty dt\, e^{-xt}/t$ is the exponential integral, and
\begin{equation}
\label{T-final}
\mathcal{T} \simeq\mathcal{S}\,\frac{1\!-\!p}{p}\int_0^\infty \!\! d\theta\,
V_\theta\int_0^\theta \!\! d u \, u \sum_{j=0}^\infty\frac{4}{(2j\!+\!1)^2}\left\{
1-e^{-(2j+1)^2/u^2}\left[1+\left(\frac{2j\!+\!1}{u}\right)^2\right]\right\}+ \mathcal{S}\,.
\end{equation}
\end{subequations}
We emphasize that $V_\theta$ defined in Eq.~\eqref{V} depends on $p$, so that
the lifetime $\mathcal{T}$ for the greedy forager does \emph{not} merely
equal the lifetime for the non-greedy forager times $(1-p)/p$.  The above
prediction for $\mathcal{T}$ agrees with our numerical simulations when
$\mathcal{S}$ is large (Fig.~\ref{greed-prelim}).  It is worth mentioning,
however that numerical evaluation of \eqref{T-final}, along with \eqref{V},
is time consuming (multiple nested integrals and sums) and unstable, so that
simulation results were more expeditious to obtain in the regime
$\mathcal{S}\leq 10^6$.

\subsection*{The negative greed regime $p\ll 1/\sqrt{\mathcal{S}}$,
  $G\approx -1$:}

In this regime, the hyperbolic function in Eq.~\eqref{finite1} can be
replaced by its argument, so that ${\widetilde F}_L(s)$ simplifies to
\begin{equation}
{\widetilde F}_L(s)\simeq \left(1+\frac{L}{2pD}s\right)^{-1}\,,
\end{equation}
while again $\widetilde{\mathcal{F}}_L(s) = {\widetilde F}_L(s)/s$.
Inverting the above Laplace transform gives
\begin{equation}
\label{Fregime2}
{\cal F}_L({\cal S})\simeq 1-e^{-2pD{\cal S}/L}\,,
\end{equation}
while Eq.~\eqref{LT} leads to $\tilde\tau_L(s)\simeq L/(2pDs)$.  Taking the
inverse Laplace transform of this quantity immediately gives
\begin{equation}
\label{first}
\tau_L(\mathcal{S})\simeq\int_{0}^{\mathcal{S}}tF_L(t)\, dt \simeq\frac{L}{2pD}\,.
\end{equation}
 
\begin{figure}[ht]
  \centerline{
\subfigure[]{\includegraphics[width=0.5\textwidth]{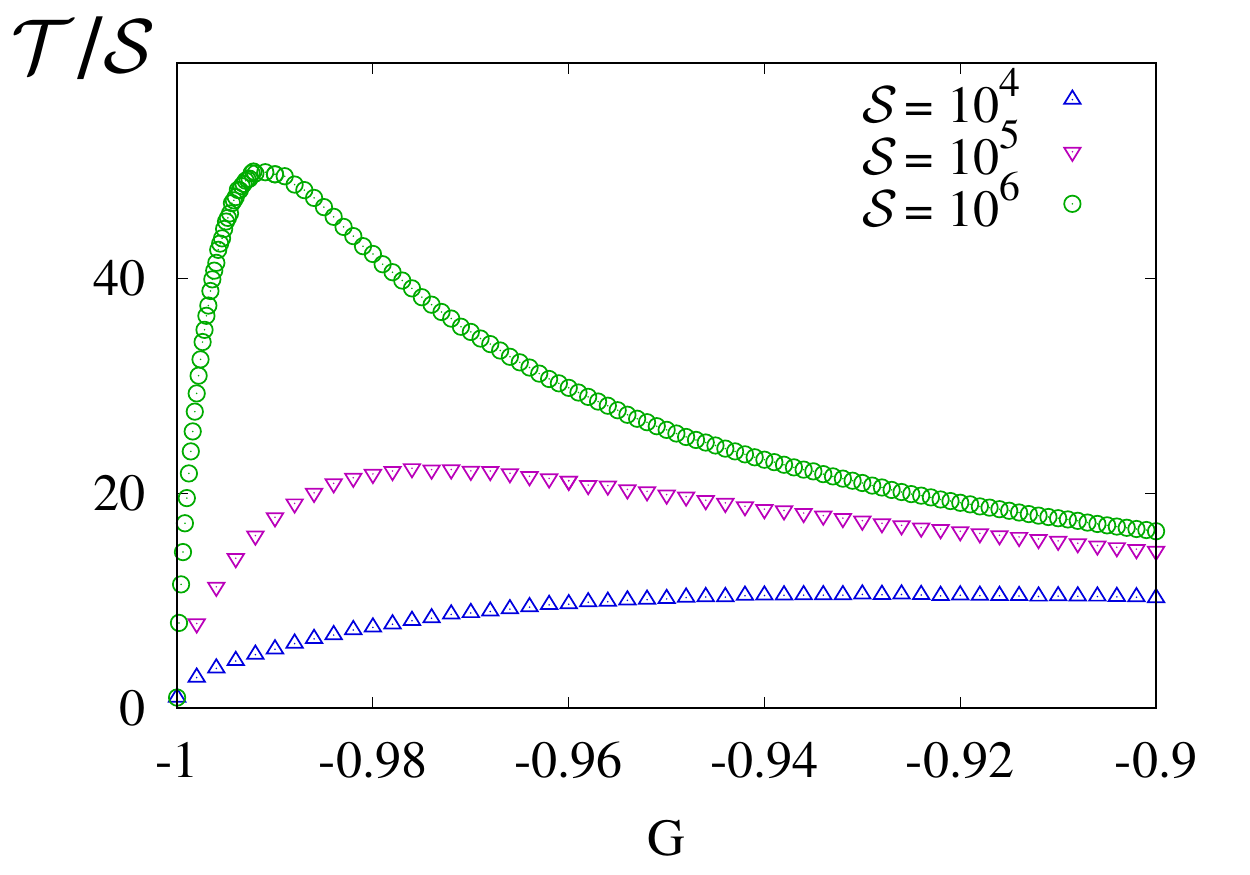}}
\subfigure[]{\includegraphics[width=0.5\textwidth]{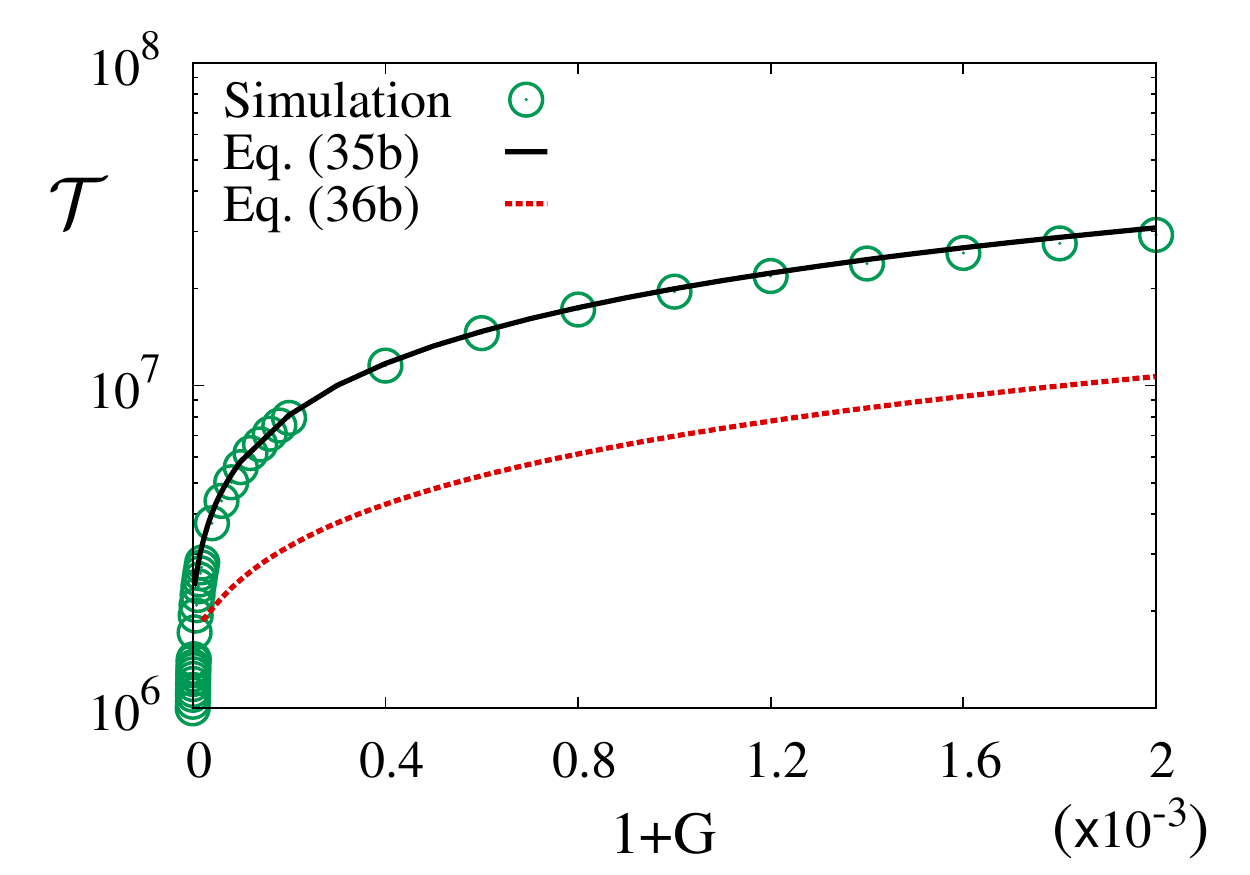}}}
\caption{Simulation data for the dependence of the scaled forager lifetime
  $\mathcal{T}/\mathcal{S}$ on $G=2p-1$ in the negative greed regime for (a)
  $G$ in the range $[-1,-0.9]$ and (b) $[-1,-0.998]$ for the case
  $\mathcal{S}=10^6$.  Also shown in (b) the full expression from
  Eq.~\eqref{tav-asymp} (solid curve), and the asymptotic result
  \eqref{tav-final} (dashed curve).}
\label{all-greed}
\end{figure}

Using the above expression for ${\cal F}_L$ and $\tau_L$, we find, for
${\cal N}$ and ${\cal T}$ (see \ref{app:NT}):
\begin{subequations}
\begin{align}
\label{N-int}
\mathcal{N}&\simeq  \sum_{n=1}^\infty n\, e^{-{2pD{\cal S}}/{n}}\,
\exp\left[-\int_1^n e^{-{2pD\mathcal{S}}/{k}}\,dk\right]\,,\\
\label{T-int}
\mathcal{T}&\simeq  \frac{1}{2p\cal S} \sum_{n=1}^\infty n^2\, e^{-{2pD{\cal S}}/{n}}\,
\exp\left[-\int_1^n e^{-{2pD{\cal S}}/{k}}\,dk\right].
\end{align}
\end{subequations}
The asymptotic evaluations of these sums and integrals are performed in
\ref{app:neg-greed-1d}, and the final results are
\begin{subequations}
\begin{align}
\label{nav-final}
\mathcal{N}&\simeq 
 \frac{\sqrt{2\pi}}{e}\,\frac{p\,\mathcal{S}}{\ln(p\,\mathcal{S})}\,,\\
\label{tav-final}
\mathcal{T}  &\simeq 
 \sqrt{\frac{\pi}{2}}\,  \frac{1}{e}\, \frac{p\,\mathcal{S}^2}{\big[
  \ln(p\,\mathcal{S})\big]^2}+\mathcal{S}\,.
\end{align}
\end{subequations}
These expressions agree with the naive estimates \eqref{N_qualit2} and
\eqref{T_qualit2} up to logarithmic corrections.  The comparison between the
asymptotic prediction for $\mathcal{T}$ in Eq.~\eqref{tav-final}, the
complete form \eqref{tav-asymp}, and simulations is shown in
Fig.~\ref{all-greed}(b).  The asymptotic approximation becomes increasingly
accurate, albeit very slowly, as $\mathcal{S}$ increases.  To observe the
true asymptotic behavior of greedy forager trajectories by simulations, it
would be necessary to treat random walks with $\mathcal{S}\gg 10^6$, a range
that is practically inaccessible.

\section{Greedy Foraging in Two Dimensions}
\label{sec:2d}

\begin{figure}[ht]
\centerline{\includegraphics[width=0.6\textwidth]{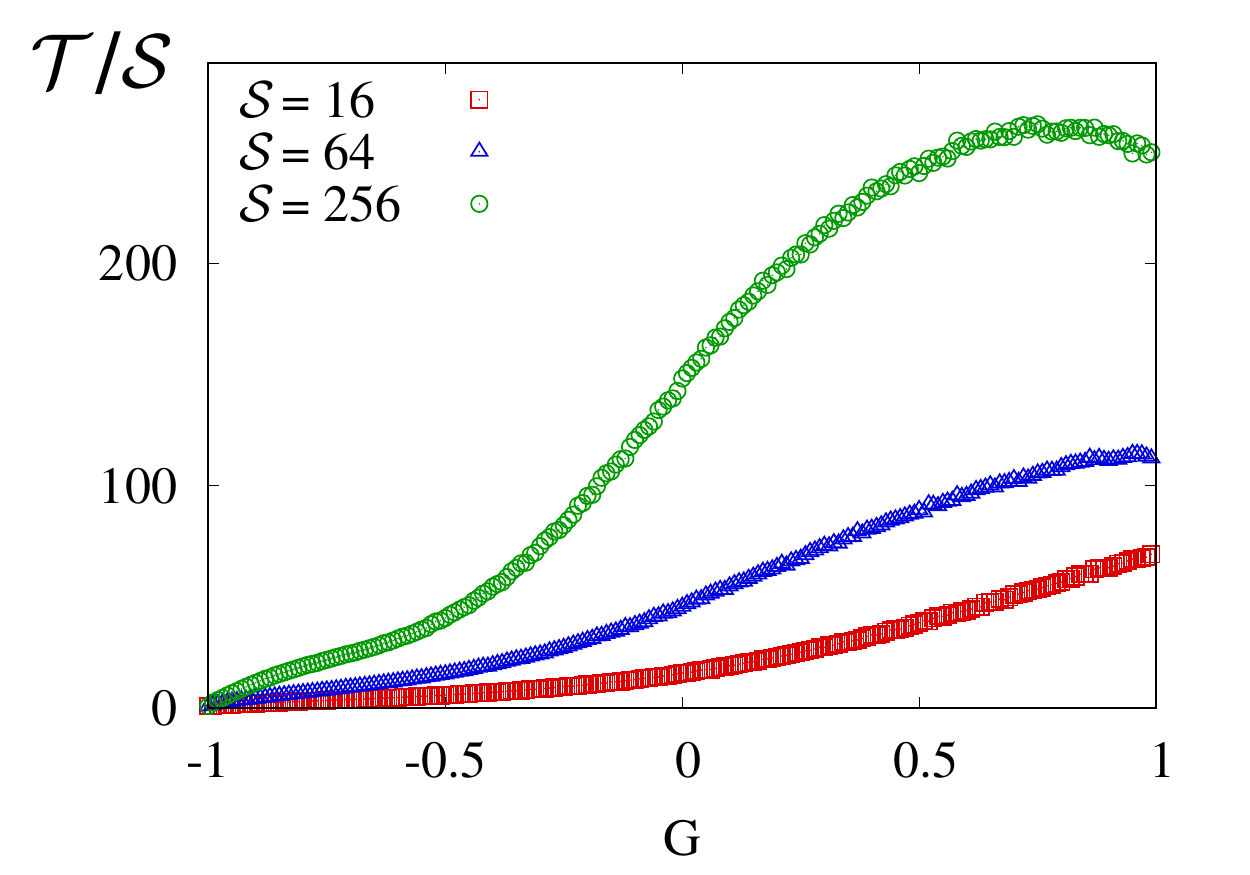}}
\caption{Average forager lifetime $\mathcal{T}$ versus greediness $G$ in two
  dimensions.  The survival times have been scaled by $\mathcal{S}$ so that
  they all fit on the same plot.}
\label{t-2d}
\end{figure}

In the dimensions, the desert carved out by a long-lived starving random
walker is typically quite ramified~\cite{BR14,CBR16}.  This geometric
complexity seems to preclude an analytic solution for the forager lifetime.
Instead, we present simulations and heuristic arguments to assess the
influence of greed on the forager lifetime.  Surprisingly, the role of
positive greed in two dimensions is opposite to that in one dimension.  For
$\mathcal{S}>\mathcal{S}^*$, with $\mathcal{S}^*\approx 90$, the lifetime
again varies non-monotonically with greediness (Fig.~\ref{t-2d}), but with
the \emph{opposite sense} to the non monotonicity compared to one dimension.
An additional perplexing feature, at first sight, is that a perfectly greedy
forager has a \emph{smaller} lifetime than a forager that is not as
avaricious.

\begin{figure}[ht]
  \centerline{\includegraphics[width=0.65\textwidth]{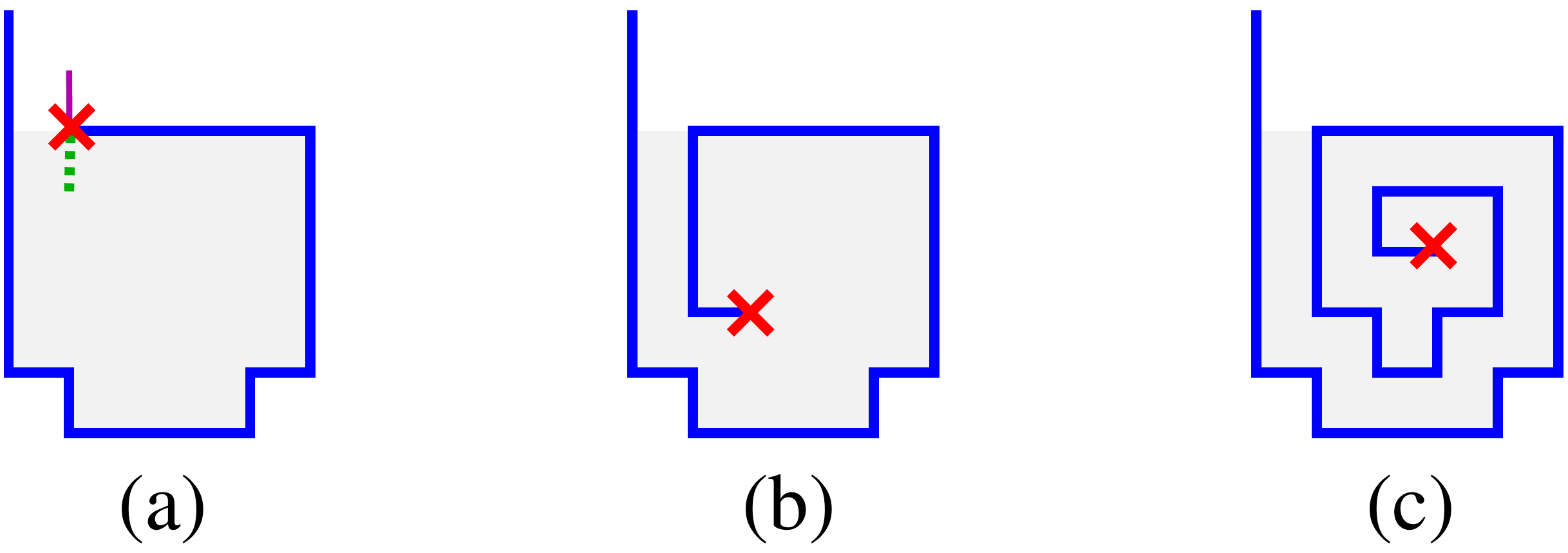}}
  \caption{ A random-walk trajectory that leads to trapping of a perfectly
    greedy forager. (a) Forager ($\times$) at the decision point.  (b)
    Forager hops to the interior region (shaded).  (c) Food in the interior
    is completely consumed, so that the forager ($\times$) may be trapped
    inside the newly created desert.}
\label{moat}
\end{figure}

We can justify this latter feature by appealing to the recurrence of a random
walk in two dimensions~\cite{F68,W94}.  Thus there will be many points where
the forager trajectory intersects itself, leading to closed loops.  Suppose
that a perfectly greedy forager is about to form a closed loop on the square
lattice, as illustrated in Fig.~\ref{moat}(a).  At this point, the forager
has only two possible choices for its next step, both of which lead to food
being consumed at the next step.  One of these choices leads to the outside
of the incipient closed loop and the other leads inside.  If the latter
choice is made, the forager is effectively self trapped by the ``moat'' that
has been created by the previous trajectory.

Once inside the moat, a perfectly greedy forager will always consume food in
its nearest neighborhood.  This consumption is interrupted when all the
current neighbors of the forager are empty.  When this happens, either the
interior is mostly or completely depleted (the latter is shown in
Fig.~\ref{moat}(c)).  While the former case is more likely, the remaining
food will be scarce and isolated.  Thus the forager carves out and then
becomes trapped inside a (perhaps slightly imperfect) desert.  In this
circumstance, the forager is likely to starve before it can escape.

Conversely, if the greediness is close to but less than 1, a forager has a
non-zero probability to cross the moat whenever it is encountered and thereby
reach food on the outside.  This mechanism provides a way for the forager to
escape the desert and survive longer than if it remained inside.  This
argument suggests that the forager lifetime must be a decreasing function of
$G$ as $G\to 1$, as confirmed by simulations (Fig.~\ref{t-2d}).  Also in
stark contrast to one dimension, there is no anomalous peak in the forager
lifetime for negative greed, at least for $\mathcal{S}\leq 256$.


\section{Discussion}

We investigated the dynamics of a greedy forager that either moves
preferentially towards food or away from food within its nearest
neighborhood.  Such a myopic greediness (or anti-greediness) represents a
particularly simple mechanism by which the motion of a forager is affected by
its environment.  In spite of its naivet\'e, the greedy forager model exhibits
rich, unexpected phenomenology that offer theoretical challenges.

As in the starving random walk without greed, a greedy forager depletes its
environment by consumption, and one basic issue is to determine the lifetime
of the forager as a function of its metabolic capacity $\mathcal{S}$ and its
greediness $G$, or equivalently, $p$, the bias that the forager experiences
when at the edge of the desert.  We determined the forager lifetime exactly
in the semi-infinite one-dimensional geometry.  Here, the lifetime grows
monotonically with greed, as might be expected naively, but the dependence of
the lifetime on $\mathcal{S}$ and $p$ is non trivial.  For sufficiently
strong greed, the forager lifetime scales as
$\mathcal{T}\sim \sqrt{\mathcal{S}}/(1-p)$, while for weak greed the lifetime
is linear in $\mathcal{S}$.

In the finite desert geometry, we found the unexpected feature that the
forager lifetime depends non-monotonically on greediness when the capacity of
the forager is sufficiently large.  Moreover, the sense of the
non-monotonicity is different in one and two dimensions.  In one dimension, a
little greed is ``bad'', while a lot of greed is ``good'', where ``bad'' and
``good'' mean decreased and increased lifetime, respectively.  We gave a
heuristic argument based on simple first-passage ideas to understand this
non-monotonicity.  In two dimensions, the opposite occurs, as a little greed
is ``good'', while being very greedy is ``bad''.  We can understand the
latter case in a simple way in terms of the self trapping of a forager.

We generalized the first-passage approach of~\cite{BR14,CBR16} to derive an
analytic expression for the average forager lifetime in one dimension that
applies in the limit of large $\mathcal{S}$.  This approach shows that a
small amount of greed is indeed detrimental for the lifetime of the forager
in one dimension.  Finally, we studied the intriguing case where a forager
has negative greed, which means that it avoids food in its local
neighborhood.  Strikingly, the lifetime of a forager in one dimension
exhibits a huge maximum when the greediness $G$ is close to $-1$, or
equivalently, $p\to 0$.  Using first-passage ideas, we argued that the
maximum lifetime occurs at a value of $p$ that scales as $\mathcal{S}^{-1/2}$
and that the lifetime at this maximum scales as $\mathcal{S}^{3/2}$.

There are several open questions about the greedy forager model that deserve
further study.  First, what is dynamics of a greedy forager in two
dimensions?  Why is the dependence of lifetime on greed opposite to that of
one dimension?  Why is there no peak in the lifetime for a forager with very
negative greed?  What is the nature of the desert geometry for different
values of greediness?  Second, what is the lifetime of a greedy forager in
greater than two dimensions?  Simulations are of limited value because the
lifetime is extremely long for non-negligible greed and memory and/or
computation time constraints become prohibitive.  What simulations can say is
obvious---in going from no greed to a specified positive value of greediness,
the increase in the forager lifetime is much greater in three dimensions than
in two.  In high dimensions, the mean-field argument given in
Refs.~\cite{BR14,CBR16} still seems to apply; this predicts that the forager
lifetime will grow as $e^{\mathcal{S}}$.  In the limit of perfect greed, it
is always possible to construct analogs of the two-dimensional cul-de-sac of
Fig.~\ref{moat}, in which a forager can enter, get trapped, and subsequently
starve.  Thus the question of whether the forager lifetime depends
non-monotonically on greed in greater than two dimensions is still open.

\section*{Acknowledgments}

OB acknowledges support from the European Research Council starting grant
No.\ FPTOpt-277998.  SR acknowledges support from grants DMR16-08211 and
DMR-1623243 from the National Science Foundation and by a grant from the John
Templeton Foundation.  UB acknowledges support from grant DMR-162324 and
grant 2012145 from the United States Israel Binational Science Foundation.

\appendix

\section{Semi-Infinite Geometry}
\label{app:semi-inf}

For fixed $p$, the large-$\mathcal{S}$ behavior of the generating function
$\widetilde{F}(z)$ in Eq.~\eqref{ginfty} may be conveniently computed in
terms of the generating function for $\mathcal{F}$ in
$\mathcal{F}(\mathcal{S})=\sum_{t=0}^\mathcal{S} F(t)$.  The relation between
these two generating functions is~\cite{AWH13}
\begin{equation}
\widetilde{\mathcal{F}}(z)=\sum_{\mathcal{S}=0}^\infty \mathcal{F}(\mathcal{S}) z^{\mathcal{S}}=\frac{{\widetilde g}(F)}{1-z}\,.
\end{equation}
Differentiating with respect to $z$ gives
\begin{align}
\label{A:F}
  \sum_{\mathcal{S}=0}^\infty \mathcal{S} \mathcal{F}(\mathcal{S})  z^{\mathcal{S}}
&=z\frac{d}{dz}\left(\frac{{\widetilde F}(z)}{1-z}\right)\nonumber\\
&\equi{z\to1}\frac{1}{(1-z)^2}-\frac{1}{\sqrt{2}}\frac{1-p}{p}\frac{1}{(1-z)^{3/2}}-\frac{1}{(1-z)}+\dots
\end{align}
We finally obtain the large-$\mathcal{S}$ behavior of
$\mathcal{F}(\mathcal{S})$ by using a discrete Tauberian
theorem~\cite{W94,H49} to give
\begin{subequations}
\begin{equation}
\label{Finfty}
\mathcal{F}(\mathcal{S})=1-\frac{1-p}{p}\sqrt{\frac{2}{\pi\mathcal{S}}}-\frac{1}{\mathcal{S}}+\dots
\end{equation}
For $p\gg 1/\sqrt{\mathcal{S}}$, we use the above expression for $\mathcal{F}(\mathcal{S})$
in Eq.~\eqref{N} to give the result quoted in \eqref{eq_N}.

When $p\to 0$ is taken before $\mathcal{S}\to\infty$, we use the following
limiting expression of ${\widetilde F}(z)$ from Eq.~\eqref{ginfty},
\begin{equation*}
\label{g2}
\lim_{p\to0}\frac{{\widetilde F}(z)}{p}=\frac{z}{\sqrt{1-z^2}},
\end{equation*}
in \eqref{A:F} to give
\begin{align*}
\lim_{p\to0}\,\,\frac{\sum_{\mathcal{S}=0}^\infty \mathcal{S}
  \mathcal{F}(\mathcal{S}) z^{\mathcal{S}}}{p}
\equi{z\to1}\frac{1}{2\sqrt{2}}\,\frac{1}{(1-z)^{5/2}}\,,
\end{align*}
so that 
\begin{align}
\label{F2}
\lim_{p\to0}\,\,\frac{\mathcal{F}(\mathcal{S})}{p}
\equi{\mathcal{S}\to\infty}\sqrt{\frac{2\mathcal{S}}{\pi}}\,.
\end{align}
\end{subequations}
Substituting the above results for $\mathcal{F}(\mathcal{S})$ in
$\mathcal{N}_{\rm
  SI}=\mathcal{F}(\mathcal{S})/\big[1-\mathcal{F}(\mathcal{S})\big]$,
from Eq.~\eqref{N}, gives
\begin{equation}
\label{Np0}
\lim_{p\to0}\,\frac{\mathcal{N}_{\rm SI}}{p}
\simeq\sqrt{\frac{\pi\mathcal{S}}{2}}\,,\qquad \mathcal{S}\to \infty\,.
\end{equation}
Note that the result for $\mathcal{N}_{\rm SI}$ quoted in Eq.~\eqref{eq_N}
holds for both $p\gg1/\sqrt{\mathcal{S}}$ and $p\ll1/\sqrt{\mathcal{S}}$.

We now determine the escape time $\tau$ for $\mathcal{S}\to\infty$,
\begin{equation}
\label{defA}
\tau= \frac{\sum_{t=0}^{\mathcal{S}}tF(t)}{\sum_{t=0}^{\mathcal{S}}F(t)}
= \frac{\sum_{t=0}^{\mathcal{S}}tF(t)}{\mathcal{F}(\mathcal{S})}
\equiv\frac{\mathcal{F}^{(1)}(\mathcal{S})}{\mathcal{F}(\mathcal{S})}\,.
\end{equation}
Following the same reasoning as given above, the generating function for
$\mathcal{F}^{(1)}(\mathcal{S})$ is
\begin{equation}
\label{diffG}
\sum_{\mathcal{S}=0}^{\infty}\mathcal{F}^{(1)}(\mathcal{S})z^{\mathcal{S}}
=\frac{z}{1-z}\,{\widetilde F} \,'(z)\,,
\end{equation}
which leads to, as $z\to1$,
\begin{equation}
  \sum_{\mathcal{S}=0}^{\infty}\mathcal{F}^{(1)}(\mathcal{S})z^{\mathcal{S}}=\frac{1}{\sqrt{2}}\frac{1-p}{p}\frac{1}{(1-z)^{3/2}}-\frac{p^2-4p+2}{p^2}\frac{1}{1-z}+\dots
\end{equation}
Using again a Tauberian theorem, the large-$\mathcal{S}$ behavior of $A$ is
\begin{equation*}
\mathcal{F}^{(1)}(\mathcal{S})=\frac{2}{\sqrt{\pi}}\frac{1-p}{p}\sqrt{\mathcal{S}}-\frac{(p^2-4p+2)}{p^2}+\cdots
\end{equation*}
so that
\begin{equation}
\tau=\frac{2}{\sqrt{\pi}}\frac{1-p}{p}\sqrt{\mathcal{S}}-\frac{(p^2-4p+2)}{p^2}+\cdots
\end{equation}
where we have also used Eqs.~\eqref{Finfty} and \eqref{defA}.  Note that the
subleading term in this expansion becomes important when $p\to1$.
Substituting the expression for $\tau$ in
$\mathcal{T}_{\rm SI} =\tau \mathcal{N}_{\rm SI} + \mathcal{S}$ gives the
final result for the lifetime quoted in Eq.~\eqref{Tregime1}.

If the limit $p\to0$ is taken before the large-$\mathcal{S}$ limit, the
expression Eq.~\eqref{g2} for ${\widetilde F}$ has to be used in
Eq.~\eqref{diffG}.  This leads to, for $p\to 0$,
\begin{equation*}
\frac{\sum_{\mathcal{S}=0}^{\infty}\mathcal{F}^{(1)}(\mathcal{S})z^{\mathcal{S}}}{p}
\equi{z\to1}\frac{3\sqrt{2}}{4(1-z)^{5/2}},
\end{equation*}
and thus
\begin{equation*}
\frac{\mathcal{F}^{(1)}(\mathcal{S})}{p}\equi{\mathcal{S}\to\infty}
\sqrt{\frac{2}{9\pi}}\,\mathcal{S}^{3/2}.
\end{equation*}
Using now Eqs.~\eqref{F2} and \eqref{defA}, we find
\begin{equation}
\tau\!\!\equi{\mathcal{S}\to \infty} \frac{1}{3}\,\mathcal{S}\,,
\end{equation}
which ultimately leads to Eq.~\eqref{Tregime1}.

\section{Escape From An Interval}
\label{sec:escape}

\begin{figure}[ht]
\centerline{\includegraphics[width=0.6\textwidth]{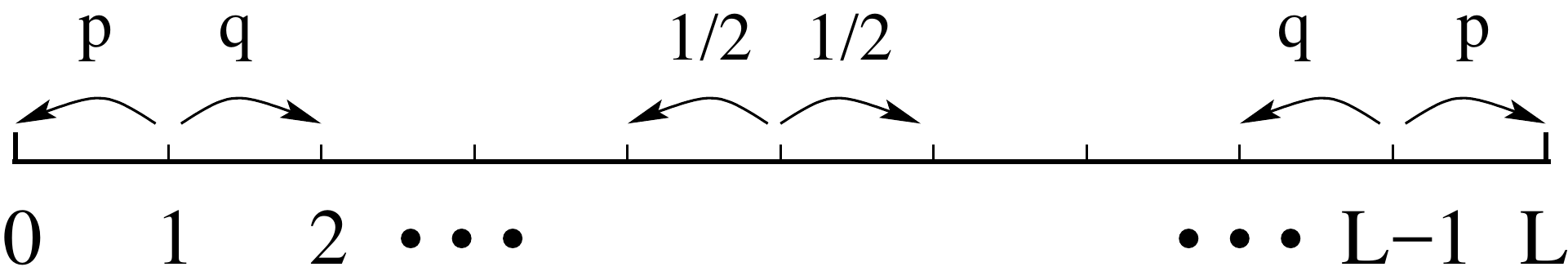}}
\caption{ Hopping probabilities for a greedy forager inside a desert of
  length $L$.}
\label{interval}
\end{figure}

We determine the first-passage properties of a random walk in a finite
interval of length $L$ whose hopping rules are the same as that of a greedy
forager.  That is, a walk in the interior hops equiprobably to the left and
right, while a walk at either $x=1$ or $x=L-1$ hops to the edge of the
interval with probability $p$ and into the interior with with probability
$1-p$ (Fig.~\ref{interval}).  For this walk, we calculate the exit
probabilities to each side of the interval, the time to exit either side of
the interval, and the conditional exit time to exit by each edge of the
interval.  We use this information in Sec.~\ref{subsec:heuristic} to argue
that the lifetime of a forager with a sufficiently large capacity varies
non-monotonically with greediness.

Let $E_n$ be the probability that the forager, which starts at site $n$,
exits the interval via the left edge.  The exit probabilities satisfy the
backward equations
\begin{align}
\begin{split}
&E_1= p +q E_2\,, \\
&E_n = \tfrac{1}{2} E_{n-1}+\tfrac{1}{2} E_{n+1} \qquad\qquad 2\leq n\leq L-2\,,\\
&E_{L-1} = qE_{L-2}\,.
\end{split}
\end{align}
No boundary conditions are needed, as the distinct equations for $n=1$ and
$n=L-1$ fully determine the exit probabilities.  As we shall see, $E_n=0$ not
at $n=L$, but at different value of $n$, and similarly for the point where
$E_n=1$.  This same behavior could be recovered by imposing the radiation
boundary condition at $n=0,L$~\cite{W94}, but this procedure involves subtleties that
are tangential to the point of the current derivation. 

Since the deviation to random-walk motion occurs only at the boundaries, we
attempt a solution that has the random-walk form in the interior of the
interval: $E_n=A+Bn$.  This ansatz automatically solves the interior
equations ($2\leq n\leq L-2$), while the boundary equations for $n=1$ and
$n=L-1$ give
\begin{align*}
E_1=p+qE_2 &~~~\longrightarrow~~~ A+B=p+q(A+2B)\,,\\
E_{L-1}=qE_{L-2} &~~~\longrightarrow~~~ A+B(L-1)=q\big(A+B(L-2)\big)\,,
\end{align*}
from which $A$ and $B$ are
\begin{equation*}
A=\frac{p(L-2)+1}{pL+2(1-2p)}\,,\qquad\qquad B=-\frac{p}{pL+2(1-2p)}\,.
\end{equation*}
Thus the probability that a greedy random walk that starts at $x=n$ exits via
the left edge of the interval is
\begin{equation}
\label{En}
  E_n=A+Bn= \frac{L-n+\frac{1}{p}(1-2p)}{L+\frac{2}{p}(1-2p)}\,,
\end{equation}
while the exit probability via the right edge is $1-E_n$.  As might be
expected for a perturbation that applies only at the boundary, the overall
effect of greed on the exit probability is small: the exit probability
changes from $E_n=1-\frac{n}{L}$ for $p=\frac{1}{2}$ to
$E_n=1-\frac{n-1}{L-2}$ for $p=1$.  That is, the effective interval length
changes from $L$ to $L-2$ as $p$ increases from $\frac{1}{2}$ to 1.


Similarly, let $t_n$ be the average time for a greedy random walker to reach
either edge of the interval when the walk starts at site $n$.  These exit
times satisfy the backward equations
\begin{align}
\begin{split}
\label{tn}
&t_1= p +q(t_2+1)\,, \\
&t_n = \tfrac{1}{2} t_{n-1}+\tfrac{1}{2} t_{n+1} +1\qquad\qquad 2\leq n\leq L-2\,,\\
&t_{L-1} = p+q(t_{L-2}+1)\,.
\end{split}
\end{align}
Again, no boundary conditions are needed, as the equations for $n=1$ and
$n=L-1$ are sufficient to solve \eqref{tn}.  We attempt a solution for these
second-order equations that has the same form as in the case of no greed:
$t_n= a+bn+cn^2$.  Substituting this ansatz into \eqref{tn} immediately gives
$c=-1$, while the equations for $t_1$ and $t_{L-1}$ lead to the conditions
\begin{align}
&-1+a+b = q(-4+a+2b)+1\,,\nonumber \\
&-(L-1)^2+b(L-1)+a = q\big[-(L-2)^2+b(L-2)+a\big]+1\,.\nonumber
\end{align}

Solving these equations, the average exit time to either edge of the interval
when starting from site $n$ is
\begin{equation}
\label{tn:app}
t_n=n(L-n)-\frac{2p-1}{p}(L-2)\,.
\end{equation}
This gives a parabolic dependence of $t_n$ on $n$ that is shifted slightly
downward compared to the case of no greed, as $p$ ranges from $\frac{1}{2}$
to 1.  Notice again that $t_n=0$ not at $n=0$ and $n=L$, but rather at points
between $n=0$ and 1 and between $n=L-1$ and $L$ for $p>\frac{1}{2}$.  This
overall shift leads to a tiny change in each $t_n$, \emph{except} when the
forager starts one site away from the boundary.  In this case, the average
exit time reduces to the expression given in Eq.~\eqref{t1}.  

Finally, and for completeness, we determine the \emph{conditional} exit
times, $t^\pm_n$, defined as the time to reach left edge of the interval when
starting from site $n$ (for $t^-$) and to the right edge (for $t^+$),
conditioned on the walker exiting only by the specified edge.  We focus on
$t^-_n$, because once $t^-_n$ is determined, we can obtain $t^+_n$ via
$t^+_n=t^-_{L-n}$.  The conditional exit times $t^-_n$ satisfy
\begin{align}
\begin{split}
\label{un}
&u_1= qu_2+E_1\,, \\
&u_n = \tfrac{1}{2} u_{n-1}+\tfrac{1}{2} u_{n+1} +E_n\qquad\qquad 2\leq n\leq L-2\,,\\
&u_{L-1} = qu_{L-2}+E_{L-1}\,,
\end{split}
\end{align}
where $u_n\equiv E_nt^-_n$, with $E_n$, the exit probability to the left
edge, given by Eq.~\eqref{En}.  Because Eqs.~\eqref{un} are second-order with
an inhomogeneous term that is linear in $n$, the general solution is a cubic
polynomial: $u_n=a+bn+cn^2+dn^3$.  Substituting this form into Eq.~\eqref{un}
for $2\leq n\leq L-2$, we obtain the conditions $c=-A$ and $d=-B/3$, where
$A$ and $B$ are the coefficient of $E_n$ in Eq.~\eqref{En}.  The remaining
two coefficients are determined by solving the equations for $u_1$ and
$u_{L-1}$ and the final results for the coefficients $a,b,c,d$ in $u_n$ are:
\begin{align}
\begin{split}
\label{abcd}
a&=\frac{2 (L-2) (1-2p) \big[p^2(L-4)(L+\frac{3}{p}(1-p))+3\big]}{3 p^3 
\big[L+\frac{2}{p}(1-2p)\big]^2}\,,\\
b&=\frac{2p^2 \big[L(L^2-6L+6)+8\big]+6pL(L-3) +6L-8 p}{3 p^2 \big[L+\frac{2}{p}(1-2p)\big]^2}\,,\\
c&=-\frac{L+\frac{1}{p}(1-2p)}{L+\frac{2}{p}(1-2p)}\,, \\
d&=-\frac{1}{3}\,\frac{1}{L+\frac{2}{p}(1-2p)}\,.
\end{split}
\end{align}
The conditional exit time to the left edge is then $t_n^-=u_n/E_n$, with
$u_n=a+bn+cn^2+dn^3$, and $E_n$ given by Eq.~\eqref{En}.  We are particularly
interested in $t^-_{L-1}$, the conditional time for a walk that starts at
$x=L-1$ to reach $x=0$.  From Eqs.~\eqref{En} and \eqref{abcd}, the limiting
behavior of this crossing time for large $L$ is given by
\begin{equation}
\label{app:tx}
t^-_{L-1}\equiv t_\times\simeq \frac{2}{3}\,L^2 + \frac{4}{3}\,\frac{L}{p}\,,
\end{equation}
which is Eq.~\eqref{tmL-1}.



\section{Catalan Triangle Numbers}
\label{app:catalan}

We define the number of paths of length $2 n$ that are adjacent to food
exactly $k$ times as $A(n, k)$.  Let $C(n)$ be the ordinary Catalan numbers,
which are defined as number of paths that return to the origin in exactly
$2 n$ steps.  These are given by
\begin{equation}
 \label{catalan-ordinary}
 C(n) = \frac{1}{n+1}\binom{2 n}{n}
\end{equation}
We can write the following convolution form that relates $A(n,k)$ with the
Catalan numbers:
\begin{equation}
\label{convolve}
 A(n,k) = 
\begin{cases}
 C(n-1) & \qquad k=1 \\ 
\displaystyle{\sum_{j=0}^{n-k} C(j) A(n-j-1,k-1)} & \qquad k\ge 2 \,.
\end{cases}
\end{equation}
This relation can be justified as follows. The number of paths for which the
forager returns to the boundary exactly once, namely, at the very end of the
trajectory, is given by the ordinary Catalan number.  The number of paths
where the forager returns $k\ge 2$ times to the boundary can be split into
the number of subpaths that return for the first time to the boundary layer
at time $2 j$, times the number of subpaths that return $k-1$ times to the
boundary layer in the remaining $2(n-j-1)$ time steps.  This gives the second
line in the right-hand side of Eq.~\eqref{convolve}.  We now define the
generating functions for $C(n)$ and $A(n,k)$,
\begin{subequations}
\label{genfunction}
  \begin{align}
    F(x) &= \sum_{n=0}^\infty C(n)  x^n\,,\\
    G(x,y)& = \sum_{n=1}^\infty \sum_{k=1}^n x^n y^k A(n,k)\,.
  \end{align}
\end{subequations}
Multiplying the left-hand side of Eq.~\eqref{convolve} by $x^n y^k$ and
summing over $n$ and $k$ with the constraint that $k\le n$, we obtain
\begin{align}
\label{genfunction2}
 \sum_{n=1}^\infty \sum_{k=1}^n x^n y^k A(n,k) &= \sum_{n=1}^\infty x^n y C(n-1) + \sum_{n=2}^\infty \sum_{k=2}^n \sum_{j=0}^{n-k} x^n y^k C(j) A(n-j-1,k-1)\,.
\end{align}
Expressing the above equation in terms of the generating function itself, we
obtain $G(x,y) = x y F(x) + x y F(x) G(x,y)$, from which the solution is
\begin{align}
G(x,y) &= \frac{x y F(x)}{1 - x y F(x)}\,.\nonumber
\end{align}
Since it is known that $F(x) = 2 / (1+\sqrt{1-4 x})$~\cite{S99}, we obtain
\begin{equation}
  G(x,y) = \frac{ 2 x y}{1 + \sqrt{1-4 x} - 2 x y}
\end{equation}

To invert Eq.~\eqref{genfunction2}, we first expand it in a power series in $y$,
 \begin{align}
  G(x,y) &= \frac{ 2 x y}{1 + \sqrt{1-4 x} - 2 x y} \nonumber \\
  &= \frac{y}{(1-\sqrt{1-4 x})/ 2 x -y} \nonumber \\
  &= \sum_{k=1}^\infty y^k \left[ \frac{2 x}{1-\sqrt{1-4 x}} \right]^k \nonumber \\
  &= \sum_{k=1}^\infty y^k \left[ 1 - \sum_{n=0}^\infty C(n) x^{n+1} \right]^k\,.
\end{align}
Numerically expanding the above expression in a double power series in $x$
and $y$ using Mathematica, we obtain
\begin{equation}
 A(n,k) = \frac{(2 n -k -1)! k}{(n-k)! n!}\,.
\end{equation}
It is worth noting that the coefficients $A(n,k)$ are related to the
\emph{Catalan Triangle Numbers} \cite{B96,LO16} $C(n,k)$ by the variable
change, $n \to n-1$, $k\to n-k$.

\section{Distribution of $\mathcal{N}$ and $\mathcal{T}$}
\label{app:NT}

We may generally write the escape probability in a finite interval of length
$L$ in the form (see Eq.~\eqref{calFk} and \eqref{Fregime2})
\begin{equation}
\mathcal{F}_L(\mathcal{S})=1-a(L,\mathcal{S})\,,
\end{equation}
with $a(L,\mathcal{S})\ll1$.  We determine the behavior of $\prod_{k=2}^n \mathcal{F}_k(\mathcal{S})$
for large $n$ by considering
\begin{equation}
\ln \prod_{k=2}^n \mathcal{F}_k(\mathcal{S}) = \sum_{k=2}^n\ln \big[1-a(k)\big]\simeq  -\sum_{k=2}^n a(k)\,.
\end{equation}
Thus $V_n$ defined in Eq.~\eqref{general_V} becomes
\begin{equation}
V_n\simeq a(n)\exp\Big[-\sum_{k=2}^n a(k)\Big]\,.
\end{equation}

We separately consider the regimes $p\gg1/\sqrt{\mathcal{S}}$ and
$p\ll1/\sqrt{\mathcal{S}}$.  For the former:
\begin{align}
\sum_{k=2}^n a(k)&=\frac{4(1-p)}{p}\sum_{k=2}^n\,\frac{1}{k}\,\,
\sum_{j=0}^{\infty}e^{-D\mathcal{S}[\pi(2j+1)/k]^2},\nonumber\\
&\simeq \frac{4(1-p)}{p}\int_{0}^n\frac{dk}{k}\,\,
\sum_{j=0}^{\infty}e^{-D\mathcal{S}[\pi(2j+1)/k]^2},\nonumber\\
&\simeq \frac{4(1-p)}{p}\int_{0}^{n/\sqrt{\mathcal{S}}}\frac{dk}{k}\,\,
\sum_{j=0}^{\infty}e^{-D[\pi(2j+1)/k]^2},\nonumber\\
&\simeq \frac{2(1-p)}{p}\sum_{n\ge0}E_1\big((2n+1)^2/\theta^2\big)\,,
\end{align}
where $\theta\equiv n/(\pi \sqrt{D\mathcal{S}})$ and
$E_1(x)\equiv\int_1^\infty {\rm d}t e^{-xt}/t$ is the exponential integral.
The distribution of the scaled variable $\theta$ is thus
\begin{equation}
\label{V}
 V_\theta\simeq \frac{4(1-p)}{p\theta}\sum_{j\ge0}e^{-(2j+1)^2/\theta^2}\,
\exp\left[-\,\frac{2(1-p)}{p}\sum_{\ell\ge0}E_1\left(\frac{(2\ell+1)^2}{\theta^2}\right)\right]\,.
\end{equation}
The average amount of food consumed by the forager at the instant of
starvation is the first moment of this distribution (see \eqref{general_V})
and immediately leads to Eq.~\eqref{N-final}.  

For $p\ll1/\sqrt{\mathcal{S}}$, we may write
\begin{equation}
\sum_{k=2}^n a(k) \simeq\int_1^n e^{-{2pD{\cal S}}/k}\,dk,
\end{equation}
which leads to
\begin{equation}
\label{Vn}
V_n\simeq e^{-2pD\mathcal{S}/n}\, \exp\left[-\int_1^n e^{-2pD\mathcal{S}/k}\right] dk\,.
\end{equation}
Substituting this result in Eq.~\eqref{general_T} immediately gives Eq.~\eqref{tav-final}.

\section{$\mathcal{N}$ and $\mathcal{T}$ for Extreme Negative Greed in 1d}
\label{app:neg-greed-1d}

We now specialize first-passage quantities to the case of extreme negative
greed; that is, $G\to -1$ or equivalently, $p\to 0$.  Here the enumeration of
trajectories to determine the distribution of food consumed at the instant of
starvation ($V_n$ in Eq.~\eqref{general_V}) greatly simplifies because the
forager typically moves back to the interior whenever it comes to the edge of
the desert.  Consider the initial condition
$\cdots\bullet\times\bullet\cdots$, which corresponds to the forager being
placed at the origin and immediately eating the food at this site.  Here
$\times$ denotes the forager and $\bullet$ a food-containing site.  Now
consider the configuration $\bullet\bullet\circ\times\bullet\bullet\cdots$
immediately after the forager has eaten a second time; here $\circ$ denotes
an empty site.  For the forager to never eat again, it must necessarily
bounce back and forth between the two empty sites $\mathcal{S}+1$ times.
Because the forager is always at the interface between food and an empty
site, each step occurs with probability $1-p$.  Thus the probability that it
does not eat again before starving, which is the same as the probability that
the forager eats exactly twice, is
\begin{equation*}
 V_2= (1-p)^{\mathcal{S}+1}\,,
\end{equation*}
while the probability for the forager to eat \emph{more than} twice is
\begin{equation*}
  \mathcal{F}_2(\mathcal{S}) =1-  (1-p)^{\mathcal{S}+1}\,.
\end{equation*}
Generally, $\mathcal{F}_k(\mathcal{S})$ was also defined as the probability
that the forager eats before it starves in an interval of $k$ empty sites
(see \eqref{general_V}).  Because the forager has already eaten $k$ times to
create this interval, it must eat more than $k$ times to escape this
interval.

Suppose now that the forager has eaten a third time.  The configuration
immediately afterward is
$\cdots\bullet\bullet\circ\circ\times\bullet\bullet\cdots$.  For the forager
to not eat again, it must next hop to the center of the interval, which
occurs with probability $(1-p)$, after which the next step necessarily takes
the forager back to the edge.  To avoid eating, the forager must repeat this
pattern of hopping to center and then back to the edge $\mathcal{S}+1$ times.
The probability for such a path of $t$ steps is $(1-p)^{t/2}$.  Thus starting
from $\cdots\bullet\bullet\circ\circ\times\bullet\bullet\cdots$, the forager
starves without again eating with probability $(1-p)^{(\mathcal{S}+1)/2}$.
Consequently, the probability that the forager eats exactly three times is
\begin{equation*}
 V_3=\mathcal{F}_2(\mathcal{S}) \,(1-p)^{(\mathcal{S}+1)/2} \,,
\end{equation*}
while the probability that the forager eats more than three times is
\begin{equation*}
  \mathcal{F}_2(\mathcal{S})\mathcal{F}_3(\mathcal{S}) 
=\big[1- (1-p)^{\mathcal{S}+1}\big] \big[1-  (1-p)^{(\mathcal{S}+1)/2}\big]\,.
\end{equation*}

Immediately after eating the fourth time, the configuration is
$\cdots\bullet\bullet\circ\circ\circ\times\bullet\bullet\cdots$.  To not eat
again, the forager must next hop away from the edge, which occurs with
probability $(1-p)$.  From the resulting state
$\cdots\bullet\bullet\circ\circ\times\circ\bullet\bullet\cdots$, the mean
time to reach either site at the edge of the desert equals 2
(\ref{sec:escape}).  For $t\gg 1$, the forager will be at the edge of the
desert $t/3$ times, on average, so that the probability that the forager will
starve without eating again is $(1-p)^{(\mathcal{S}+1)/3}$.  Thus the
probability that the forager eats exactly four times before starving
is\footnote{The factor $(\mathcal{S}+1)/k$ that appears in the exponent
  should be modified by even-odd oscillations that will arise when
  $\mathcal{S}$ is either even or odd.  Since we are interested in the limit
  of large $\mathcal{S}$, there even-odd effects should play a small role
  asymptotically, and we ignore them.}
\begin{equation*}
  V_4= \mathcal{F}_2(\mathcal{S})\mathcal{F}_3(\mathcal{S}) \,(1-p)^{(\mathcal{S}+1)/3}\,.
\end{equation*}
Continuing this reasoning, the probability that the forager eats more than
$n-1$ times before starving is (after changing variables from $\mathcal{S}$ to
$z$)
\begin{subequations}
\begin{equation}
\label{Qn-prod}
 \prod_{k=2}^{n-1} \mathcal{F}_k(z) = 
 \prod_{k=1}^{n-2} (1- z^{1/k})\,,\qquad\qquad z\equiv (1-p)^{\mathcal{S}+1}\,,
\end{equation}
where we shifted the index in the second product.  Thus the probability for
the forager to eat exactly $n$ times before starving is
\begin{equation}
V_n=  \prod_{k=2}^{n-1}\mathcal{F}_k(z) \,z^{1/(n-1)}\,.
\end{equation}
\end{subequations}

To compute $\mathcal{N}$ and $\mathcal{T}$ in Eqs.~\eqref{N-final} and
\eqref{T-final}, we first investigate the nature of the function
$\prod\mathcal{F}_n(z)$ in \eqref{Qn-prod}.  This function equals 1 for small
$n$ and sharply drops to 0 for sufficiently large $n$.  Each term in the
product equals $(1-z^{1/k})$ and $z$ is, in general, small.  Naively, one
might think that the point where $\prod\mathcal{F}_n$ crosses over from being
nearly 1 to decaying would occur when $z^{1/n}=\frac{1}{2}$, or
$n=\ln z/\ln\frac{1}{2}$.  Numerically, however, $\prod\mathcal{F}_n$ is
already vanishingly small at this value of $n$ because each term in the
product is only slightly different than its immediate predecessor.  At the
point where a term in the product is close to $\frac{1}{2}$, the product of
all previous terms is already close to zero because there are many preceding
terms that are also close to $\frac{1}{2}$.

An important consequence of this location of the crossover, is that
$z^{1/k}\equiv e^{-a/k}$ is small over the entire range where
$\prod\mathcal{F}_n$ is non zero.  Thus we may simplify $\prod\mathcal{F}_n$
by the following standard manipulations:
\begin{align}
\label{lnQ}
\ln\prod_{k=2}^{n-1}\mathcal{F}_k & = \ln\Big[\prod_{k=1}^{n-2} (1- e^{-a/k})\Big]\,, \nonumber\\
&= \sum_{k=1}^{n-2} \,\ln (1- e^{-a/k})\,,\nonumber\\
&\simeq  -\int_1^n e^{-a/k}\, dk
= e^{-a}- n\,e^{-a/n}-a\big[E_1(a)-E_1(a/n)\big]\,,
\end{align}
where $E_1(\cdot)$ is again the exponential integral.  The above form is the
explicit representation of the exponential factors in Eqs.~\eqref{N-int}
and~\eqref{T-int}.  We are interested in the regime where $p$ is small, but
$p>1/\mathcal{S}$, so that a forager is likely to carve out a desert of an
appreciable size.  In this case, we have
\begin{equation}
a = -\ln(1-p)^{\mathcal{S}+1} \simeq p\mathcal{S}\,.
\end{equation}

Using the representation \eqref{lnQ} for $\prod\mathcal{F}_n$ as well as
$z=e^{-a}$, the amount of food consumed at the instant of starvation is
\begin{align}
\label{nav-asymp}
\mathcal{N} = \sum_{n\geq 0} n V_n
 &\simeq  \int_0^\infty dn\,  n \, e^{-a/n} \,
\exp\Big\{e^{-a}- n\,e^{-a/n}-a\big[E_1(a)-E_1(a/n)\big]\Big\}\,,\nonumber \\
&= \int_0^\infty   dn\,
\exp\Big\{\ln n -a/n+ e^{-a}- n\,e^{-a/n}-a\big[E_1(a)-E_1(a/n)\big]\Big\}\,,\nonumber\\
&\equiv   \int_0^\infty dn\,  \exp\big[f(n)\big]\,.
\end{align}
The function $f(n)$ in \eqref{nav-asymp} has a single peak whose width
vanishes slowly as $a\to\infty$.  Thus we evaluate this integral by the
Laplace method.  Differentiating $f(n)$ in \eqref{nav-asymp} with respect to
$n$ and setting the result to zero gives
\begin{equation}
\label{fprime}
  f'(n)= \frac{1}{n} +\frac{a}{n^2}- e^{-a/n}=0\,.
\end{equation}
Numerically, we find that $f$ is maximized at a value $n^*$ that grows
slightly slower than linearly with $a$.  The first natural hypothesis
$n^*=a/\ln a$, fails to make the terms in \eqref{fprime} balance.  Thus we
attempt a solution of the form $n^*\simeq a(1+\epsilon)/\ln a$, where
$\epsilon\ll 1$ for large $a$, and look for a self-consistent solution for
$\epsilon$.  Substituting in \eqref{fprime}, we find that the leading
behaviors of the second two terms in this equation are dominant and they
balance when
\begin{equation}
\label{nstar}
n^*\simeq \frac{a}{\ln a}\Big[1+2\,\frac{\ln \ln a}{\ln a}\Big]\,.
\end{equation}

To complete the evaluation of the integral \eqref{nav-asymp}, we also need
the second derivative of $f$ evaluated at $n^*$.  To leading order, this is
\begin{align}
\label{fpp}
f''(n^*)\simeq -\frac{1}{(n^*)^2} -\frac{2a}{(n^*)^3}-\frac{a}{(n^*)^2}\, e^{-a/n^*}
\simeq - \frac{(\ln a)^4}{a^2}\,,
\end{align}
where the dominant contribution comes only from the term
$(a/n^2)\, e^{-a/n}$.  Finally, we need $f^*\equiv f(n^*)$.  Here, we need to
keep the first two terms in the asymptotic expansion of $E_1$ to obtain,
after some straightforward algebra
\begin{align}
f^* &=  \ln n^* -\frac{a}{n^*} - \frac{(n^*)^2}{a}\, e^{-a/n^*} 
 \simeq \ln\ln a -1\,.
\end{align}
Assembling all these elements gives
\begin{align}
\label{N-app}
\mathcal{N} &\simeq  \int_0^\infty dn\,
                           \exp\big[f(n)\big] 
\sim   \sqrt{\frac{2\pi}{|f''|}}\,\,  e^{f^*}
\simeq  \sqrt{\frac{\pi}{2}}\,\frac{a^2}{e} (\ln a)^{-2}\,,\nonumber\\
&\sim  \frac{\sqrt{2\pi}}{e}\,\frac{p\,\mathcal{S}}{\big[ \ln(p\,\mathcal{S})\big]}\,,
\end{align}
which is Eq.~\eqref{nav-final} in the text.

Similarly, the average forager lifetime is
\begin{align}
\label{tav-asymp}
  \mathcal{T} &\simeq \sum_{n\geq 0} \,\frac{n^2}{2p}\, V_n
\simeq  \frac{1}{2p} \int_0^\infty dn\,  \exp\big[g(n)\big]\,,
\end{align}
with $g(n)\equiv f(n)+\ln n$, and $f(n)$ given in \eqref{nav-asymp}.
Following the same steps as in the evaluation of $\mathcal{N}$, we find the
same maximizing value of $n^*$, and the same width of the peak,
$g''(n^*)=f''(n^*)$, while $g(n^*)$ now equals $\ln a -1$.  Assembling everything gives
\begin{align}
\label{T-app}
\mathcal{T} &\simeq  \frac{1}{2p} \int_0^\infty dn\,
                           \exp\big[f(n)\big] +\mathcal{S}
\sim  \frac{1}{2p} \sqrt{\frac{2\pi}{|f''|}}\,\,  e^{f^*}+\mathcal{S}
\simeq \frac{1}{p}\, \sqrt{\frac{\pi}{2}}\,\frac{a^2}{e} (\ln a)^{-2}+\mathcal{S}\,,\nonumber\\
&\sim \sqrt{\frac{\pi}{2}}\,\frac{1}{e}\,\frac{p\,\mathcal{S}^2}
{\big[\ln(p\,\mathcal{S})\big]^2}+\mathcal{S}\,.
\end{align}
which is Eq.~\eqref{tav-final} in the text.  

We should note some caveats about these calculations for $\mathcal{N}$ and
$\mathcal{T}$.  Normally in applying the Laplace method, the contribution
from the peak of the distribution, $e^f$, is exponentially larger than the
contribution of the width of the maximum, $1/\sqrt{|f''|}$.  This is not the
case here, as the width contribution is almost of the same magnitude as that
of the peak (for $\mathcal{T}$) or larger than the peak contribution (for
$\mathcal{N}$).  In addition, the integrals for $\mathcal{N}$ and
$\mathcal{T}$ in Eqs.~\eqref{nav-asymp} and \eqref{tav-asymp}, respectively,
have peaks that are \emph{movable}.  To recast such integrals into a form
where the Laplace method can be applied, one normally introduces a rescaled
coordinate so as to fix the location of the maximum~\cite{BO78}.  If the
function in the exponent is algebraic, this rescaling is trivial.  However,
it does not seem possible to implement such a rescaling for our functions
$f(n)$ and $g(n)$.  Thus our results \eqref{N-app} and \eqref{T-app} have to
be viewed with some suspicion.  We checked, however, that the result
\eqref{T-app} moves closer to the exact integral \eqref{tav-asymp} as
$\mathcal{S}$ increased far beyond the values that we can simulate, but the
convergence is extremely slow.


\newpage

\begin{thebibliography}{99}

\bibitem{C76} E. L. Charnov, Theor.\ Popul.\ Biol.\ {\bf 9}, 129 (1976).

\bibitem{KR85} P. Knoppien and J. Reddingius, J. Theor.\ Biol.\ {\bf 114},
  273 (1985).

\bibitem{SK86} D. W. Stephens and J. R. Krebs, {\it Foraging Theory},
  (Princeton University Press, Princeton, NJ, 1986).

\bibitem{OB90} W. J. O'Brien, H. I. Browman, and B. I. Evans, Am.\ Sci.\ {\bf
    78}, 152 (1990).

\bibitem{B91} J. W. Bell, Searching Behaviour, the Behavioural Ecology of
  Finding Resources, Animal Behaviour Series (Chapman and Hall, London,
  1991).

\bibitem{ASD97} J. P. Anderson, D. W. Stephens, and S. R. Dunbar, Behav.\
  Ecol.\ {\bf 8}, 307 (1997).

\bibitem{KM01} L. D. Kramer and R. L. McLaughlin, Am.\ Zool.\ {\bf 41}, 137 (2001).

\bibitem{Oaten:1977} A. Oaten, Theor.\ Popul.\ Biol.\ \textbf{12}, 263
  (1977).

\bibitem{Iwasa:1981} Y. Iwasa, M. Higashi, and N. Yamamura, Amer.\
  Nat.\textbf{117}, 710 (1981).

\bibitem{Green:1984} R. F. Green, Amer.\ Nat.\ \textbf{123}, 30 (1984).

\bibitem{Viswanathan:1999a} G. Viswanathan, S.~V. Buldyrev, S. Havlin,
  M. Da~Luz, E. Raposo, and H.~E. Stanley, Nature \textbf{401}, 911 (1999).

\bibitem{Benichou:2005} O. B\'enichou, M. Coppey, M. Moreau, P.-H. Suet, and
  R. Voituriez, Phys.\ Rev.\ Lett.\ \textbf{94}, 198101 (2005).

\bibitem{Oshanin:2007} G. Oshanin, H. Wio, K. Lindenberg, and S. Burlatsky,
  J. Phys.\ Condens.\ Matter \textbf{19}, 065142 (2007).

\bibitem{Lomholt:2008} M.~A. Lomholt, K. Tal, R. Metzler, and K. Joseph,
  Proc.\ Natl.\ Acad.\ Sci.\ (USA) \textbf{105}, 11055 (2008).

\bibitem{Bressloff:2011} P.~C. Bressloff and J.~M. Newby, Phys.\ Rev.\ E
\textbf{83}, 061139 (2011).

\bibitem{Tejedor:2012} V. Tejedor, R. Voituriez, and O. B{\'e}nichou, Phys.\
  Rev.\ Lett.\ \textbf{108}, 088103 (2012).





\bibitem{BR14} O. B\'enichou and S. Redner, Phys.\ Rev.\ Lett.\ {\bf 113},
  238101 (2014).

\bibitem{CBR16} M. Chupeau, O. B\'enichou, and S. Redner, J. Phys.\ A: Math.\
  \& Theor.\ {\bf 49}, 394003 (2016).

\bibitem{BRB17} U. Bhat, S. Redner, and O. B\'enichou, Phys.\ Rev.\ E {\bf
    96}, 062119 (2017).

\bibitem{BB72} H. C. Berg and D. A. Brown, Nature \textbf{239}, 500
  (1972).

\bibitem{BP77} H. C. Berg and E. M. Purcell, Biophys.\ J. \textbf{20}, 193 (1977).

\bibitem{DZ88} P. N. Devreotes and S. H. Zigmond, Annu.\ Rev.\ Cell Biol.\
  \textbf{4}, 649 (1988).

\bibitem{EB02} E. Balkovsky and B. I. Shraiman, Proc.\ Natl.\ Acad.\ Sci.\
  (USA) {\bf 99}, 12589 (2002).

\bibitem{infotaxis} M. Vergassola, E. Villermaux, and B. I. Shraiman, Nature
  \textbf{445}, 406 (2007).

\bibitem{G05} R. Grima, Phys.\ Rev.\ Lett.\ {\bf 95}, 128103 (2005).

\bibitem{TKMI16} L. Tweedy, D. A. Knecht, G. M. Mackay, and R. H. Insall,
  PLoS Biol.\ {\bf 14}, e1002404 (2016).

\bibitem{JKM17} C. Jin, C. Kruger, and C. C. Maass, Proc.\ Natl.\ Acad.\
  Sci.\ (USA) {\bf 114}, 5099 (2017).


\bibitem{F68} W. Feller, {\it Introduction to Probability Theory and its
    Applications, $3^{\rm rd}$ ed.}, (J. Wiley \& Sons, New York, 1968)

\bibitem{W94} G. H. Weiss, {\it Aspects and Application of the Random Walk},
  (North-Holland, Amsterdam, 1994).

\bibitem{R01} S. Redner, {\it A Guide to First-Passage Processes}, (Cambridge
  University Press, Cambridge, UK, 2001).

\bibitem{AWH13} G. B. Arfken, H. J. Weber, and F. E. Harris,
  \emph{Mathematical Methods for Physicists: A Comprehensive Guide,
    $7^{\rm th}$ ed.}, (Academic Press, Waltham, MA, 2013).

\bibitem{H49} G. H. Hardy, \emph{Divergent Series}, (Oxford University Press,
  Oxford, UK, 1949).

\bibitem{B96} D. F. Bailey, Math.\ Mag.\ {\bf 69}, 128 (1996).

\bibitem{LO16} K.-H. Lee and S.-J. Oh, arXiv.org: 1601.06685.


\bibitem{S99} R. P. Stanley, {\it Enumerative Combinatorics, Vol.\ 2},
  (Cambridge University Press, Cambridge, UK, 1999).

\bibitem{BO78} C. M. Bender and S. O. Orszag, {\it Advanced Mathematical
    Methods for Scientists and Engineers}, (McGraw-Hill, New York, 1978).

\end{thebibliography}
\end{document}